%% file: articlerevise2007039.tex
\documentclass[12pt]{article}
\usepackage{graphicx,xspace} 
\graphicspath{{Eps/}}
\usepackage[english,francais]{babel}
\usepackage[T1]{fontenc}
\usepackage{ae,aecompl} 

\small\normalsize 
\title{Approche variationnelle pour le calcul bayésien dans les problèmes inverses en imagerie}

\author{Ali Mohammad-Djafari \\ 
Laboratoire des signaux et systèmes \\
(UMR 08506, CNRS-SUPELEC-Univ Paris Sud) \\ 
Supélec, Plateau de Moulon, 91192 Gif-sur-Yvette Cedex, France}

\date{\today}

\input{alphabet.tex}
\input{macros1.tex}

\def\nub{\bm{\nu}}
\def\etat{\widetilde{\eta}}
\def\nubt{\widetilde{\nub}}
\def\ubt{\widetilde{\ub}}
\def\phibt{\widetilde{\phib}}

\begin{document}
\bcc
{\Huge Approche variationnelle pour le calcul bayésien dans les problèmes inverses en imagerie}
\ecc

\bcc
{\large
Ali Mohammad-Djafari \\ 
Laboratoire des signaux et systèmes \\
(UMR 08506, CNRS-SUPELEC-Univ Paris Sud) \\ 
Supélec, Plateau de Moulon, 91192 Gif-sur-Yvette Cedex, France
}
\\ ~ \\ \today \\ 
\ecc

\begin{abstract}
{Dans une approche bayésienne non supervisée pour la résolution d'un problème 
inverse, on cherche à estimer conjointement la grandeur d'intérêt $\fb$ et 
les paramètres $\thetab$ à partir des données observées $\gb$ et un modèle 
$\Mc$ liant ces grandeurs. Ceci se fait en utilisant la loi \apost conjointe 
$p(\fb,\thetab|\gb;\Mc)$. 
L'expression de cette loi est souvent complexe 
et son exploration et le calcul des estimateurs bayésiens nécessitent soit 
les outils d'optimisation 
de critères ou de calcul d'espérances des lois multivariées. 
Dans tous ces cas, il y a souvent besoin de faire des approximations. 
L'approximation de Laplace et les méthodes d'échantillonnage MCMC sont deux 
approches classiques 
(analytique et numérique) qui ont été explorés avec succès pour ce fin. 
Ici, nous étudions l'approximation de 
$p(\fb,\thetab|\gb)$ par une loi séparable en $\fb$ et en $\thetab$. Ceci permet 
de proposer des algorithmes itératifs plus abordables en coût de calcul, surtout, 
lorsqu'on choisit ces lois approchantes dans des familles des lois exponentielles conjuguées. 
Le principal objet de ce papier est de présenter les différents algorithmes 
que l'on obtient pour différents choix de ces familles. \`A titre d'illustration, 
nous considérons le cas de la restauration d'image par déconvolution simple ou 
myope avec des \aprio séparables, markoviens simples ou avec des champs cachés.}
\end{abstract}

\newpage
\bcc
{\Huge Variational Approche for Bayesien Computation for Inverse Problems in Imaging Systems}
\ecc

\bcc
{\large
Ali Mohammad-Djafari \\ 
Laboratoire des signaux et systèmes \\
(UMR 08506, CNRS-SUPELEC-Univ Paris Sud) \\ 
Supélec, Plateau de Moulon, 91192 Gif-sur-Yvette Cedex, France
}
\\ ~ \\ \today \\ 
\ecc

\bcc
Abstract
\ecc

{In a non supervised Bayesian estimation approach for inverse problems in imaging systems, 
one tries to estimate jointly the unknown image pixels $\fb$ and the hyperparameters $\thetab$ 
given the observed data $\gb$ and a model $\Mc$ linking these quantities. 
This is, in general, done through the joint posterior law $p(\fb,\thetab|\gb;\Mc)$. 
The expression of this joint law is often very complex and its 
exploration through sampling and computation of the point estimators such as 
MAP and posterior means need either optimization of or integration of multivariate 
probability laws. In any of these cases, we need to do approximations. 
Laplace approximation and sampling by MCMC are two approximation methods, respectively analytical 
and numerical, which have been used before with success for this task. 
In this paper, we explore the possibility of approximating this joint law by a separable one in $\fb$ 
and in $\thetab$. This gives the possibility of developing iterative algorithms with more reasonable 
computational cost, in particular, if the approximating laws are choosed in the exponential conjugate 
families. The main objective of this paper is to give details of different algorithms we obtain with 
different choices of these families. To illustrate more in detail this approach, we consider the case of image restoration by simple or myopic deconvolution with separable, simple markovian or hidden markovian models.}

\newpage
\section{Introduction}

Une présentation simplifiée et synthétique des problèmes inverses en imagerie, en se plaçant en dimensions finies, consiste à vouloir retrouver une grandeur inconnue $\fb$ à partir des observations $\gb$ d'une grandeur observée, en supposant connaître un modèle $\Mc$ qui les lient. 
La forme la plus simple de ce modèle est un modèle linéaire de la forme 
\beq
\Mc : \qquad \gb=\Hb \fb + \epsilonb
\label{eq01}
\eeq
où on suppose que toutes les erreurs de modélisation et de mesure peuvent être représentées par $\epsilonb$. Notons aussi que $\gb$ et $\fb$ sont, en général, des vecteurs de grandes dimensions, ce qui signifie que nous considérons ici le cas discrétisé où $\gb$ contient l'ensemble des grandeurs mesurées et $\fb$ l'ensemble des valeurs qui décrivent la grandeur inconnue. Dans ce contexte $\Hb$ est une matrice dont les éléments sont définies par le modèle et les étapes de discrétisation du problème. 

Dans une approche estimation bayésienne non supervisée pour résoudre un problème inverse, 
d'abord on utilise ce modèle pour définir la loi de probabilité $p(\gb|\fb,\thetab_1;\Mc)$ où $\thetab_1$ représente l'ensemble des paramètres qui décrivent cette loi. Lors que cette fonction est considérée comme une fonction de $\fb$ et de $\thetab_1$, elle est appelée la vraisemblance des inconnues $\fb$ et $\thetab$ du modèle $\Mc$. Son expression s'obtient à partir de la loi de probabilité des erreurs $\epsilonb$ en utilisant le modèle (\ref{eq01}).  
Par exemple, lorsque $\epsilonb$ dans ce modèle est modélisé par un vecteur aléatoire centré, 
blanc, gaussien et de covariance fixée $\Sigmabe=\sigmae^2\Ib$, on a 
\beq
p(\gb|\fb,\thetab_1;\Mc)=\Nc(\Hb\fb, \Sigmabe)
\label{eq02}
\eeq
où $\thetab_1=\sigmae^2$. D'autres lois avec d'autres paramètres $\thetab_1$ peuvent bien sûr être utilisées. 

La deuxième étape dans cette approche est l'attribution ou le choix d'une loi dite \aprio $p(\fb|\thetab_2;\Mc)$ pour les inconnues $\fb$, où $\thetab_2$ représente ses paramètres. 
La troisième étape consiste à écrire l'expression de la loi \apost des inconnues $\fb$~:
\beq
p(\fb|\thetab,\gb;\Mc)=
\frac{p(\gb,\fb|\thetab;\Mc)}{p(\gb|\thetab;\Mc)}=
\frac{p(\gb|\fb,\thetab_1;\Mc) \; p(\fb|\thetab_2;\Mc)}{p(\gb|\thetab;\Mc)}, 
\label{eq03}
\eeq
où on suppose implicitement connaître l'ensemble des paramètres $\thetab=(\thetab_1, \thetab_2)$. 
Mais, dans un cas réel, nous sommes amenés souvent à les estimer aussi. Pour cela, dans l'approche bayésienne, on leur attribue aussi une loi \aprio $p(\thetab|\Mc)$, et l'on obtient alors une loi \apost conjointe des inconnues $\fb$ et des hyperparamètres $\thetab=(\thetab_1, \thetab_2)$~:
\beqn
p(\fb,\thetab|\gb;\Mc)&=&\frac{p(\gb,\fb,\thetab|\Mc)}{p(\gb|\Mc)} \nonumber \\
&=&\frac{p(\gb|\fb,\thetab_1;\Mc) \; p(\fb|\thetab_2;\Mc)\; p(\thetab|\Mc)}{p(\gb|\Mc)}.
\label{eq04}
\eeqn

Dans cette relation, le dénominateur  
\beq
p(\gb|\Mc)= \intd p(\gb|\fb,\thetab;\Mc) \; p(\fb|\thetab;\Mc) \; p(\thetab|\Mc) \d{\fb} \d{\thetab}
\label{eq05}
\eeq
est la vraisemblance marginale du modèle $\Mc$ dont son logarithme $\ln p(\gb|\Mc)$ est appellé \emph{évidence du modèle} $\Mc$. 

Afin d'introduire les notions qui vont être utilisées dans la suite de ce travail, il est intéressant 
de mentionner que, pour n'importe quelle loi de 
probabilité $q(\fb,\thetab)$ (dont nous verrons le choix et l'utilité par la suite), l'évidence du 
modèle $\ln p(\gb|\Mc)$ vérifie 
\beqn
\ln p(\gb|\Mc)
&=& \ln \intd p(\gb,\fb,\thetab|\Mc) \d{\fb} \d{\thetab} \nonumber \\ 
&=& \ln \intd q(\fb,\thetab) \frac{p(\gb,\fb,\thetab|\Mc)}{q(\fb,\thetab)} \d{\fb} \d{\thetab} \nonumber \\ 
&\ge& \intd q(\fb,\thetab) \ln \frac{p(\gb,\fb,\thetab|\Mc)}{q(\fb,\thetab)} \d{\fb} \d{\thetab}.
\label{eq06}
\eeqn
(d'après l'inégalité de Jensen~: $\ln(\esp{p/q}) \ge \esp{\ln(p/q)}$). 
Aussi, notant par  
\beq
\Fc(q)=\intd q(\fb,\thetab) \ln \frac{p(\gb,\fb,\thetab|\Mc)}{q(\fb,\thetab)} \d{\fb} \d{\thetab}
\label{eq07}
\eeq
et par 
\beq
\mbox{KL}(q:p)=\intd q(\fb,\thetab) \ln \frac{q(\fb,\thetab)}{p(\fb,\thetab|\gb;\Mc)} \d{\fb} \d{\thetab}
\label{eq08}
\eeq 
on montre facilement (en remplaçant $p(\gb,\fb,\thetab|\Mc)=p(\fb,\thetab|\gb;\Mc)\; p(\gb|\Mc)$ dans l'expression de $\Fc(q)$) que
\beq
\ln p(\gb|\Mc)=\Fc(q)+\mbox{KL}(q:p).
\label{eq09}
\eeq 
Ainsi $\Fc(q)$, appelée l'énergie libre de $q(\fb,\thetab)$ par rapport à $p(\gb,\fb,\thetab|\Mc)$, est une limite inférieure de 
$\ln p(\gb|\Mc)$ car $\mbox{KL}(q:p)\ge 0$. 
Par la suite, nous allons écrire l'expression de $\Fc(q)$ par 
\beq 
\Fc(q)= \left<\ln{p(\gb,\fb,\thetab;\Mc)}\right>_q+\Hc(q)
\label{eq10}
\eeq
où nous avons utilisé la notation $<.>_q$ pour l'espérance suivant la loi $q$ et $\Hc(q)$ 
est l'entropie de $q$~:
\beq
H(q)= - \intd q(\fb,\thetab) \ln {q(\fb,\thetab)} \d{\fb} \d{\thetab}
\label{eq11}
\eeq 

Arrivé à ce stade, les questions posées sont~:
\bit
\item {\bf Inférence~:} \'Etant donnée les expressions des lois \apost (\ref{eq03}) et (\ref{eq04}), 
comment les utiliser pour définir une solution au problème inverse décrit en (\ref{eq01}) ? 
\item {\bf Sélection de modèle~:} Comment peut-on sélectionner un modèle parmi un ensemble de 
modèles $\Mc_i$. 
\eit
En ce qui concerne le problème de l'inférence, les principaux choix sont les estimateurs au sens 
du Maximum \apost (MAP) ou au sens de la moyenne \apost (PM). Dans le premier cas, on a besoin 
des outils d'optimisation et dans le deuxième cas des outils d'intégration (analytique ou numérique). 
Pour la sélection du modèle, nous nous contenterons ici de noter que l'expression de la 
vraisemblance du moèle (\ref{eq05}) peut être utilisés à cette fin. 
Dans ce papier, nous nous focalisons sur la première question où on cherche à inférer $\fb$ et $\thetab$ utilisant la loi \apost jointe (\ref{eq04}). 

Pour les problèmes inverses, la solution au sens du MAP a été utilisée avec succès pour sa 
simplicité et en raison de son lien avec l'approche déterministe de la régularisation. 
Mais, il y a des situations où cette solution ne donne pas satisfaction, et où la solution 
au sens de la moyenne \apost peut être préférée. Mais, les situations où on peut avoir une 
solution analytique pour les intégrations qui sont nécessaires pour obtenir ces estimées sont rares. 
Il y a alors pratiquement deux voies~: \\ 

\noindent{\bf Intégration numérique par échantillonnage~:} Il s'agit d'approcher les espérances par 
des moyennes empiriques des échantillons générés suivant la loi \apost. Toute la difficulté est 
alors de générer ces échantillons, et c'est là qu'interviennent les méthodes de MCMC 
(Markov Chain Monté Carlo). Le principal intérêt de ces méthodes est qu'elles permettent 
d'explorer l'ensemble de l'espace de la loi \apost, mais l'inconvénient majeur est leur coût 
de calcul qui est dû au nombre important d'itérations nécessaire pour la convergence des 
chaînes et le nombre important de points qu'il faut générer pour obtenir des estimations de 
bonnes qualités.  

\noindent{\bf Approximation de la loi \apost par des lois plus simples~:} Il s'agit de reporter le calcul des intégrales après une simplification par approximation de la loi \apost. Une première approximation utilisée historiquement est \emph{Approximation de Laplace} où on approxime la lois \apost par une loi gaussienne. Dans ce cas, les deux estimateurs MAP et PM sont équivalents et tous les calculs deviennent analytiques.

Une deuxième solution est d'approcher la \apost par une loi séparable, ce qui permet de réduire la dimension des intégrations. 
Cette voie est plus récente et le principal objet de ce papier.

De façon générale, l'idée d'approcher une loi conjointe $p(\xb)$ de plusieurs variables $\xb$ par une loi séparable 
$q(\xb)=\prod_j q_j(x_j)$ n'est pas nouveau et peut être trouvée dans la littératures de la fin des années 90:~\cite{Rustagi76,Ghahramani97,Penny1998,Roberts1998,Attias1999,Jordan1999,Penny1999}. 
Le choix d'un critère pour mesurer la qualité de cette approximation et l'étude des effets de cette approximation sur les qualités des estimateurs obtenus apparaît dans les travaux plus récents~\cite{Attias00,jaakkola00,Miskin00,Miskin2000,Miskin2001,Penny2002,Roberts2002,Cassidy2002,Penny2003,Penny2003a,Choudrey2003,Choudrey03a}. L'usage de cette approche en estimation des paramètres d'un modèle d'observation avec des variables cachées en statistique est également récente~\cite{will_hmm_ica,Choudrey03a,Blei04,Nasios2004,Archambeau2004,Woolrich2005,Blei06a,Beal2006,Kim2006,Wang06,Watanabie06,Nasios2006,Friston2006,wp_vb06,Ghahramani2007,McGrory2007,Forbes2007}. 
Dans la plupart de ces travaux, l'application est plutôt en classification en utilisant un modèle de mélanges. Dans le domaine du traitement du signal, ils utilisent des modèles de mélange avec des étiquettes des classes modélisées par des chaînes de Markov. Dans le domaine du traitement d'image, la plupart de ces travaux sont consacrés à la segmentation d'image en utilisant soit un modèle séparable ou markovien pour les étiquettes. Cette approche a été aussi utilisé récemment en séparation de sources \cite{Ichir2005a,Ichir2005b,Snoussi2007} et en traitement des images hyperspectrales \cite{Bali2005a,Bali2008}. 
L'application réelle de cette approche pour simplifier les calculs bayésiens dans les problèmes inverses avec un opérateur mélangeant est l'originalité de ce travail \cite{Mohammad-Djafari2007a,Ayasso2008,Mohammad-Djafari2008a}. En effet, dans la plupart de ces travaux, avec les notations utilisées dans ce papier, on suppose qu'on a observé directement $\fb$ dont la loi est modélisée par un mélange de gaussiennes et le principal objectif de ces méthodes est l'estimation des paramètres de ce mélange et la sélection du modèle. Dans d'autres travaux on considère un modèle d'observation ponctuel où $g_i=h(f_i)+\epsilon_i$ et l'objective est la segmentation de l'image $\fb$. 
Les travaux dans lesquels, on utilise l'approche variationnelle pour des problèmes inverses de restauration ou de reconstruction d'images sont assez rare. On trouve essentiellement les travaux \cite{Molina99,Likas2004,Blekas2005} qui considèrent le cas des problèmes inverses linéaires, mais soit avec des modèles \aprio de Gauss-Markov ou avec des variables cachées de contours ou d'étiquettes des régions séparables. La dépendance spatiale de ces variables cachées n'est pas prise en compte. 

La suite de ce papier est organisée de la manière suivante~: une présentation très synthétique de l'approche variationnelle est fournie dans la section 2. 
Dans la section 3, nous nous intéresserons à l'approximation de la loi \apost $p(\fb,\thetab|\gb;\Mc)$ pour les problèmes inverses. Il s'agit alors d'appliquer l'approche variationnelle pour ce cas particulier avec différent choix pour les familles des lois approchantes. 
Dans la section 4, nous détaillerons l'application de la méthode au cas de la restauration simple ou myope d'image. Enfin, dans la section 5, nous décrirons la manière d'appliquer cette méthode au cas de la restauration d'image avec une modélisation \aprio plus complexe~: un modèle hiérarchique de Gauss-Markov avec un champ caché de Potts pour des étiquettes des régions dans l'image. 
En effet, cette modélisation convient pour bien des cas de restauration et de reconstruction d'images dans des applications où l'image recherchée représente un objet composé d'un nombre fini de matériaux homogènes. C'est exactement cette information \aprio qui est modélisée par un modèle de mélange avec des étiquettes $\zb$ markoviennes \cite{Mohammad-Djafari2006e}. Ici, donc, la loi \apost $p(\fb,\zb,\thetab|\gb;\Mc)$ doit être approximée par des lois plus simples à utiliser. 
Finalement, dans la section 6, nous résumons les apports de ce papier. 
Il est à noter cependant, que dans ce papier, seules les principes des méthodes proposées sont présentés et les détails de mise en oeuvre de ces méthodes, ainsi que des résultats de simulations et évaluation des performances de ces algorithmes sont reportées à un prochaine papier.

\section{Principe de l'approche variationnelle}
Considérons le problème général de l'approximation d'une loi conjointe $p(\xb|\Mc)$ de plusieurs variables $\xb$ par une loi séparable $q(\xb)=\prod_j q_j(x_j)$. videment, il faut choisir un critère. Considérons le critère de divergence $\mbox{KL}(q:p)$ entre $p$ et $q$~: 
\[
\mbox{KL}(q:p)= \intg q(\xb) \ln \frac{q(\xb)}{p(\xb|\Mc)} \d{\xb} 
\]
et cherchons la solution $\qh(\xb)$ qui le minimise. 
La solution optimale ne peut être calculée que d'une manière itérative, car ce critère n'étant pas quadratique en $q$, sa dérivée par rapport à $q$ n'est pas linéaire et une solution explicite n'est pas disponible. On peut alors envisager une solution itérative coordonnée par coordonnée.   

Notons $q_{-j}(\xb_{-j})=\prod_{i\not=j} q_i(x_i)$, ce qui permet d'écrire $q(\xb)=q_j(x_j)\,q_{-j}(\xb_{-j})$ où $\xb_{-j}=\{x_i, i\not=j\}$ et dévéloppons 
ce critère : 
\beqn
\mbox{KL}(q:p)&=& - \intg q(\xb) \ln \frac{p(\xb|\Mc)}{q(\xb)} \d{\xb} \nonumber\\ 
              &=& - H(q) - \left< \ln p(\xb|\Mc)\right>_{q(\xb)} \nonumber\\ 
              &=& - \sum_j H(q_j) - \left< \ln p(\xb|\Mc)\right>_{q(\xb)} \nonumber\\  
              &=& - \sum_j H(q_j) - \int \left< \ln p(\xb|\Mc)\right>_{q_{-j}} q_j(x_j) \d{x_j}
\label{eq12}
\eeqn 
où $H(q)$ est l'entropie de $q$ et $H(q_j)$ est l'entropie de $q_j$.  

Notons que si $q_{-j}$ est fixée $\mbox{KL}(q:p)$ est convexe en $q_j$ et sa minimisation sous la contrainte de normalisation de $q_j$ s'écrit
\beq
q_j(x_j) = \frac{1}{C_j} \expf{\left< \ln p(\xb|\Mc)\right>_{q_{-j}}}
\label{eq13}
\eeq 
avec 
\beq
C_j=\int \expf{\left< \ln p(\xb|\Mc)\right>_{q_{-j}}} \d{x_j}
\label{eq14}
\eeq 
On note que l'expression de $q_j$ dépend de celle de ${q_{-j}}$ et que, l'obtention de $q_j$ (étant donnée $q_{-j}$) se fait aussi d'une manière itérative en deux étapes~:
\beq
\left\{\barr{ll}
\Qc(q_j)&=H(q_j) + \left<\ln p(\xb|\Mc)\right>_{q_j)},\\
q_j^{(k+1)}  &=\argmax{q_j}{\Qc(q_j)}, 
\earr\right.
\label{eq15}
\eeq
ce qui ressemble à un algorithme du type EM (Espérance-Maximisation) généralisé, car dans la première équation nous avons à calculer une espérance et dans la deuxième étape nous avons une maximisation. 

Les calculs non paramétriques sont souvent trop coûteux. 
On choisit alors une forme paramétrique pour ces lois de telle sorte que l'on puisse, à chaque itération, 
remettre à jour seulement les paramètres de ces lois, à condition cependant que ces formes ne 
changent pas au cours des itérations. La famille des lois exponentielles conjuguées ont cette propriété  
\cite{Ghahramani97,Molina99,Patriksson99,Choudrey2000,Penny2000,Choudrey03b,Nasios2006}. 

La première étape de simplification de ce calcul est donc de considérer la famille des lois paramétriques $q_j(x_j|\thetab_j)$ où $\thetab_j$ est un vecteur de paramètres. 
En effet, dans ce cas, l'algorithme itératif précédent se transforme à~: 
\beq
\left\{\barr{ll}
\Qc(\thetab_j)&=H\left(q_j(x_j|\thetab_j)\right) + \left<\ln p(\xb|\Mc)\right>_{q_j(x_j|\thetab_j)},\\
\thetab_j^{(k+1)}  &=\argmax{\thetab_j}{\Qc(\thetab_j)}, 
\earr\right.
\label{eq16}
\eeq
ce qui ressemble plus à un algorithme du type EM. 

Une deuxième étape de simplification est de choisir pour $p(\xb|\thetab;\Mc)$ la famille des lois exponentielles conjuguées~: 
\beq
p(\xb|\thetab;\Mc)=f(\xb) \, g(\thetab) \, \expf{\phib(\thetab)^T \ub(\xb)}
\label{eq17}
\eeq 
où $\phib(\thetab)$ est le vecteur de paramètres naturel et $f(\xb)$, $g(\thetab)$ et $\ub(\xb)$ sont des fonctions connues. Il est alors facile de montrer que $q_j(x_j)$ dans l'équation (\ref{eq13}) sera aussi dans la famille des lois exponentielles conjuguées, et par conséquence $q(\xb|\thetabt)$ restera dans la même famille et nous aurons juste à remettre à jours ses paramètres.   

\noindent{\bf Remarque~:} Cette famille de lois a une propriété dite \emph{conjuguée} dans le sens que si on choisit comme \aprio 
\beq
p(\thetab|\eta,\nub)=h(\eta,\nub) \, g(\thetab)^{\eta} \, \expf{\nub^T\phib(\thetab)}
\label{eq18}
\eeq 
l'\apost correspondant 
\beq
p(\thetab|\xb;\Mc) \propto p(\xb|\thetab;\Mc) \; p(\thetab|\eta,\nub)
\label{eq19}
\eeq 
sera aussi dans la même famille (\ref{eq18}). La famille des lois \aprio (\ref{eq17}) est alors dites \emph{conjuguée} de la famille des lois (\ref{eq16}).

\bigskip 
Pour cette famille de lois, on a 
\beq
\left<\ln p(\xb|\thetab;\Mc)\right>_q=\left<\ln f(\xb)\right>_q + \ln g(\thetab) + \phib(\thetab)^T \left<\ub(\xb)\right>_q
\label{eq20}
\eeq
et donc la forme de la loi séparable $q_j$ correspondante est 
\beq
q_j(x_j|\thetabt)\propto g(\thetabt) \expf{<f(\xb)>_{q_{_j}}+\phib(\thetabt)^T <\ub(\xb)>_{q_{_j}}}
\label{eq21}
\eeq
où  $\thetabt$ désigne les paramètres particuliers de $q_j$ qu'il faut mettre à jour au cours des itérations. 

Par ailleurs, sachant que $q(\xb)$ est séparable, si $\ln p(\xb|\thetab;\Mc)$ est polynomial en $\xb$, le calcul de l'espérance~(\ref{eq20}), mais surtout celui de sa dérivée par rapport aux $q_j$ et aux paramètres $\thetab$, qui sont nécessaire pour l'optimisation, seront facilités. 
Comme nous le verrons par la suite, pour les problèmes inverses, nous avons choisi ce genre de lois. 

\section{Approche variationnelle pour les problèmes inverses linéaires}
Nous allons maintenant utiliser ces relations pour décrire le principe de l'approche variationnelle au cas particulier des problèmes inverses (\ref{eq01}) où l'utilisation directe de la loi \apost conjointe $p(\fb,\thetab|\gb;\Mc)$ dans l'équation (\ref{eq04}) 
est souvent trop coûteuse pour pouvoir être explorée par échantillonnage direct de type Monté Carlo ou pour calculer les moyennes \apost 
\beq
\fbh=\int\int \fb \; {p(\fb,\thetab|\gb;\Mc)} \d{\thetab} \d{\fb}
\label{eq22}
\eeq
et
\beq
\thetabh=\int\int \thetab \; {p(\fb,\thetab|\gb;\Mc)} \d{\fb} \d{\thetab}.
\label{eq23}
\eeq
En effet, rare sont les cas où on puisse trouver d'expressions analytiques pour ces intégrales. 
De même l'exploration de cette loi par des méthodes de Monté Carlo est aussi coûteuses. 
On cherche alors à l'approcher 
par une loi plus simple $q(\fb,\thetab)$. Par simplicité, nous entendons par exemple une loi $q$ qui soit séparable en $\fb$ et en $\thetab$~:
\beq
q(\fb,\thetab)= q_1(\fb) \; q_2(\thetab)
\label{eq24}
\eeq
videmment, cette approximation doit être faite de telle sorte qu'une mesure de distance entre $q$ et $p$ soit minimale. Si, d'une manière naturelle, on choisi $\mbox{KL}(q:p)$ comme cette mesure, on aura:
\beq
(\qh_1, \qh_2)=\argmin{(q_1, q_2)}{\mbox{KL}(q_1 q_2 : p)}=\argmax{(q_1, q_2)}{\Fc(q_1 q_2)}
\label{eq25}
\eeq
et sachant que $\mbox{KL}(q_1 q_2 : p)$ est convexe en $q_1$ à $q_2$ fixée et vice versa, 
on peut obtenir la solution d'une manière itérative~: 
\beq
\left\{\barr{ll}
\qh_1&=\argmin{q_1}{\mbox{KL}(q_1 \qh_2 : p)}=\argmax{q_1}{\Fc(q_1 \qh_2)}; \\ 
\qh_2&=\argmin{q_2}{\mbox{KL}(\qh_1 q_2 : p)}=\argmax{q_2}{\Fc(\qh_1 q_2)}.
\earr\right.
\label{eq26}
\eeq
Utilisant la relation (\ref{eq07}), il est facile de montrer que les solutions d'optimisation   
de ces étapes sont 
\beq
\left\{\barr{ll}
\qh_1(\fb)&\propto \expf{\left<\ln{p(\gb,\fb,\thetab;\Mc)}\right>_{\qh_2(\thetab)}} \\ 
\qh_2(\thetab)&\propto \expf{\left<\ln{p(\gb,\fb,\thetab;\Mc)}\right>_{\qh_1(\fb)}} 
\earr\right.
\label{eq27}
\eeq
Une fois cet algorithme convergé vers $\qh^*_1(\fb)$ et $\qh^*_2(\thetab)$, on peut les utiliser 
d'une manière indépendante pour calculer, par exemple les moyennes \quad 
$\fbh^*    = \int \fb \; \qh^*_1(\fb) \d{\fb}$  \quad et \\ 
$\thetabh^*= \int \thetab \;\; \qh^*_2(\thetab) \d{\thetab}$.

Comme nous l'avons déjà indiqué, les calculs non paramétriques sont souvent trop coûteux. 
On choisit alors une forme paramétrique pour ces lois de telle sorte que l'on puisse, à chaque itération, 
remettre à jour seulement les paramètres de ces lois. Nous examinons ici, trois cas:

\subsection{Cas dégénéré} 

On prend pour $q_1(\fb)$ et $q_2(\thetab)$ des formes dégénérées suivantes~:
\beq
\left\{\barr{ll}
q_1(\fb|\fbt)&=\delta(\fb-\fbt) \\ 
q_2(\thetab|\thetabt)&=\delta(\thetab-\thetabt) 
\earr\right.
\label{eq28}
\eeq
Par conséquence, au cours des itérations, nous n'aurons qu'à remettre à jour $\fbt$ et $\thetabt$. En remplaçant $q_1(\fb)$ et $q_2(\thetab)$ dans les relations  (\ref{eq27}) on obtient~:
\beq
\left\{\barr{ll}
\qh_1(\fb)\propto p(\gb,\fb,\thetabt;\Mc) \propto p(\fb,\thetabt|\gb;\Mc)  \\ 
\qh_2(\thetab) \propto p(\gb,\fbt,\thetab;\Mc) \propto p(\fbt,\thetab|\gb;\Mc).
\earr\right.
\label{eq29}
\eeq
Il est alors facile de voir que la recherche de $\fbt$ et $\thetabt$ au cours des itérations 
devient équivalent à~:
\beq
\left\{\barr{@{}l@{}l@{}l@{}}
\fbt    &=\argmax{\fb}{p(\gb,\fb,\thetabt;\Mc)}    &=\argmax{\fb}{p(\fb,\thetabt|\gb;\Mc)} \\ 
\thetabt&=\argmax{\thetab}{p(\gb,\fbt,\thetab;\Mc)}&=\argmax{\thetab}{p(\fbt,\thetab|\gb;\Mc)}   
\earr\right.
\label{eq30}
\eeq
On remarque alors que l'on retrouve un algorithme de type MAP joint ou ICM (Iterated Conditional Mode). L'inconvénient majeur ici est qu'avec le choix (\ref{eq28}), les incertitudes liées à 
chacune des inconnues $\fb$ et $\thetab$ ne sont pas prise en compte pour l'estimation de l'autre inconnue. 

\subsection{Cas particulier conduisant à l'algorithme EM} 
On prend comme dans le cas précédent une forme dégénérée pour 
$q_2(\thetab)=\delta(\thetab-\thetabt)$, ce qui donne 
\beq
\qh_1(\fb)\propto p(\gb,\fb,\thetabt;\Mc) \propto p(\fb,\thetabt|\gb;\Mc)\propto p(\fb|\thetabt,\gb;\Mc)
\label{eq31}
\eeq
ce qui signifie que $\qh_1(\fb)$ est une loi dans la même famille que la loi \apost 
$p(\fb|\thetab,\gb;\Mc)$. 
\'Evidemment, si la forme de cette loi est simple, par exemple une gaussienne,  
(ce qui est le cas dans les situations que nous étudierons) les calculs seront simples. 

A chaque itération, on aurait alors à remettre à jour $\thetabt$ 
qui est ensuite utilisé pour trouver $\qh_1(\fb|\thetabt)=p(\fb|\thetabt,\gb;\Mc)$, elle même  utilisée pour calculer 
\beq
Q(\thetab,\thetabt)=\left<\ln{p(\gb,\fb,\thetab;\Mc)}\right>_{\qh_1(\fb|\thetabt)}.
\label{eq32}
\eeq
On remarque facilement l'équivalence avec l'algorithme EM qui se résume à~: 
\beq
\left\{\barr{@{}l@{}l@{}l@{}}
\mbox{E~:~~} Q(\thetab,\thetabt)&=\left<\ln{p(\gb,\fb,\thetab;\Mc)}\right>_{\qh_1(\fb|\thetabt)}, \\ 
\mbox{M~:~~} \thetabt&=\argmax{\thetab}{Q(\thetab,\thetabt)}.
\earr\right.
\label{eq33}
\eeq
L'inconvénient majeur ici est que l'incertitude liée à $\thetab$ n'est pas prise en compte pour l'estimation de $\fb$. 

\subsection{Choix particulier proposé pour les problèmes inverses linéaires} 

Il s'agit de choisir, pour $q_1(\fb)$ et $q_2(\thetab)$ des lois dans les mêmes familles de lois que celles de 
$p(\fb|\gb,\thetab)$ et de $p(\thetab|\gb,\fb)$. En effet, comme nous le verrons plus bas, dans le cas des problèmes inverses linéaires (\ref{eq01}) avec des choix approprié pour les lois \aprio associée à la modélisation directe du problème, ces lois \apost conditionnelles restes dans les mêmes familles, ce qui permet de profiter de la mise à jour facile 
de ces loi. 

Dans ce travail, dans un premier temps, nous allons considéré le cas des 
problèmes inverses linéaires~(\ref{eq01})~: $\gb=\Hb \; \fb + \epsilonb$,  
où $\Hb$ représente la forme discrétisé de la modélisation directe du problème et 
$\epsilonb$ représente l'ensemble des erreurs de mesure et de modélisation  avec des hypothèses suivantes:   
\beq
\barr{rcl}
p(\gb|\Hb,\fb,\theta_e;\Mc)     &=& \Nc(\Hb\fb,(1/\theta_e)\Ib), \\ 
p(\fb|\theta_f;\Mc)             &=& \Nc(\zerob,(1/\theta_f)(\Db^t_f\Db_f)^{-1}),\\ 
p(\theta_e;\Mc)                 &=& \Gc(\alpha_{e0},\beta_{e0}),\\ 
p(\theta_f;\Mc)                 &=& \Gc(\alpha_{f0},\beta_{f0}),
\earr 
\label{eq34}
\eeq
où $\Db_f$ est la matrice des différences finies d'ordre un ou deux et $\thetab=(\theta_e=1/\sigma_{\epsilon}^2,\theta_f=1/\sigma_f^2)$. 
On obtient alors facilement les expressions de $p(\gb,\fb|\thetab;\Mc)$, $p(\fb|\thetab,\gb;\Mc)$ et  $p(\thetab|\gb,\fb;\Mc)$ qui sont~:
\beq
\barr{rcl}
p(\gb,\fb|\Hb,\theta_e;\Mc) &=& \Nc(\Hb\fb,(1/\theta_e)\Ib) \times \\ 
&&  \Nc\left(\zerob,(1/\theta_f)(\Db^t_f\Db_f)^{-1}\right), \\ 
p(\fb|\gb,\Hb,\theta_f;\Mc) &=& \Nc(\mubh_f,\Sigmabh_f),\\ 
p(\theta_e|\gb;\Mc)         &=& \Gc(\alphah_{e},\betah_{e}), \\ 
p(\theta_f|\gb;\Mc)         &=& \Gc(\alphah_{f},\betah_{f})
\earr
\label{eq35}
\eeq
où les expressions de $\mubh_f$, $\Sigmabh_f$, $(\alphah_{e},\betah_{e})$ et 
$(\alphah_{f},\betah_{f})$ sont~: 
\beq
\barr{lcl}
\Sigmabh_f &=& [<\theta_e> \Hb^t \Hb+<\theta_f>\Db_f^t\Db_f]^{-1}\\ 
           &=& \frac{1}{<\theta_e>} [\Hb^t \Hb+\lambda\Db_f^t\Db_f]^{-1}, 
\mbox{~avec~} \lambda=\frac{<\theta_f>}{<\theta_e>}\\ 
\mubh_f    &=& <\theta_e> \Sigmabh_f \Hb^t \gb,
\\
\alphah_{e}               &=& \alpha_{e0} + M/2,\\ 
\betah_{e}                &=& \beta_{e0} + 1/2 <\epsilonb^t\epsilonb>,
\\
\alphah_{f}               &=& \alpha_{f0} + N/2,\\
\betah_{f}                &=& \beta_{f0} + 1/2 \, \trace{\Db_f^t\Db_f<\fb\fb^t>},  
\earr
\label{eq36}
\eeq
où
\beq
\barr{ll}
<\fb>=\mubh_f, & <\fb\fb^t>=\mubh_f\mubh_f^t + \Sigmabh_f, \\ 
               & <\epsilonb^t\epsilonb>=[\gb-\Hb\mubh_f]^t[\gb-\Hb\mubh_f], \\ 
<\theta_{e}>=\alpha_{e}/\beta_{e}, & 
<\theta_{f}>=\alpha_{f}/\beta_{f} 
\earr
\label{eq37}
\eeq
L'algorithme de mise à jour ici devient : 
\bit
\item Initialiser 
$\alphah_{e}=\alpha_{e0},\betah_{e}=\beta_{e0},
 \alphah_{f}=\alpha_{f0},\betah_{f}=\beta_{f0}$
\item Mettre à jour jusqu'à la convergence : \\ 
$<\theta_{e}>$ et $<\theta_{f}>$ puis 
$\Sigmabh_f$ et puis $\mubh_f$ \\ 
et puis $\alphah_{e}$, $\betah_{e}$, $\alphah_{f}$, $\betah_{f}$. 
\eit

La principale difficulté ici est l'inversion de la matrice $[\Hb^t \Hb+\lambda\Db_f^t\Db_f]$ 
qui est de dimensions trop grande. 
Une solution est de choisir pour $q(\fb|\gb,\thetab)$ aussi une loi séparable $q(\fb|\gb,\thetab)=\prod_j q(f_j|\gb,\thetab)$. 
L'autre solution, comme nous la verrons dans la section suivante, est d'utiliser la structure spécifique de la matrice à inverser pour trouver un algorithme 
d'inversion convenable. 

\section{Application en restauration d'image}
Dans le cas de la restauration d'image où $\Hb$ a une structure particulière, et où l'opération 
$\Hb\fb$ représente une convolution de l'image $f$ avec la réponse impulsionnelle $h$, la partie difficile et coûteuse de ces calculs est celle du calcul de $\fbh$ qui peut se faire à l'aide de la Transformée de Fourier rapide \cite{Hunt71}. 

De même, l'approche peut être étendue pour le cas de la 
restauration aveugle ou myope où on cherche à la fois d'estimer la réponse impulsionnelle $\hb$, 
l'image $\fb$ et les hyperparamètres $\thetab$. 
Pour établir l'expressions des différentes lois dans ce cas, nous notons que le problème direct, suivant que l'on s'intéresse à $\fb$ (déconvolution) ou  à $\hb$ (identification de la réponse impulsionnelle), peux 
s'écrire 
\beq 
\barr{rcl}
g(\rb)&=& h(\rb)*f(\rb)+\epsilon(\rb) =f(\rb)*h(\rb)+\epsilon(\rb) \\ 
\gb   &=& \Hb \; \fb + \epsilonb =\Fb \; \hb + \epsilonb
\earr
\label{eq38}
\eeq
où la matrice $\Hb$ est entièrement définie par le vecteur $\hb$ et la matrice $\Fb$ est entièrement définie par le vecteur $\fb$. 

Pour permettre d'obtenir une solution bayésienne pour l'étape de l'identification, nous devons 
aussi modéliser $\hb$. Une solution est de supposer $\hb=\Phib \wb$ où la matrice $\Phib$ 
est une matrice telle que $\Phib \wb$ représente la convolution $\phi(\rb)*w(\rb)$. 
Ainsi les colonnes de $\Phib$ représentent une base et les éléments du vecteur $\wb$ représentent 
les coefficients de la décomposition de $h$ sur cette base. On a ainsi
\beq 
\barr{rcl}
g(\rb)&=& (\phi*w)*f(\rb)+\epsilon(\rb) =f*(\phi*w)(\rb)+\epsilon(\rb) \\ 
\gb   &=& \Phib \; \Wb \; \fb + \epsilonb =\Fb \; \Phib \; \wb + \epsilonb
\earr
\label{eq39}
\eeq
où la matrice $\Wb$ est entièrement définie par le vecteur $\wb$. 

Le problème de la déconvolution aveugle se ramène à l'estimation de $\fb$ et $\wb$ 
avec des lois
\beq
p(\gb|\wb,\fb,\Sigmabe) = \Nc(\Phib\Wb\fb,\Sigmabe)
= \Nc(\Fb\Phib\wb,\Sigmabe)
\label{eq40}
\eeq
\mbox{~avec~} 
\[
\Sigmabe=\diag{\frac{1}{\theta_{e_i}}, i=1,\cdots,M} 
\mbox{~et~} p(\theta_{e_i})= \Gc(\alpha_{e0},\beta_{e0}),
\]
\beq 
p(\fb|\theta_f)             = \Nc\left(\zerob,(\theta_f\Db^t_f\Db_f)^{-1}\right)  
\label{eq41}
\eeq
\mbox{~avec~} 
\[
p(\theta_f)= \Gc(\alpha_{f0},\beta_{f0}),
\]
et
\beq
p(\wb|\alphab)= \prod_j p(w_j|\alpha_{w_j}) = \prod_j \Nc(0,\frac{1}{\alpha_{w_j}}) 
\label{eq42}
\eeq
\mbox{~avec~}
\[
p(\alphab) = \prod_j p(\alpha_{w_j}) = \prod_j \Gc(a_0,b_0). 
\]
Avec ces lois \aprio, il est alors facile de trouver l'expression de la loi conjointe 
$p(\fb,\wb,\Sigmabe,\theta_f,\alphab;\gb)$ et la loi \apost  $p(\fb,\wb,\Sigmabe,\theta_f,\alphab|\gb)$. 
Cependant l'expression de cette loi  
\beq
\barr{ll}
p(\fb,\wb,\Sigmabe,\theta_f,\alphab|\gb) \propto & 
p(\gb|\wb,\fb,\Sigmabe) \,  p(\fb|\theta_f) \, p(\wb|\alphab) \\ 
& p(\theta_e) \, p(\theta_f) \, p(\alphab)
\earr 
\label{eq43}
\eeq
n'est pas séparable en ses composantes. L'approche variationnelle consiste donc à l'approcher par une loi séparable 
\beq
p(\fb,\wb,\thetab_e,\theta_f,\alphab|\gb) \simeq q(\fb) q(\wb) \prod_i q(\theta_{e_i}) q(\theta_f) \prod_j q(\alpha_{w_j})
\label{eq44}
\eeq
et avec les choix des lois \aprio conjuguées en appliquant la procédure décrite plus haut, on obtient 
\beqn
q(\fb)    &=& \Nc(\mub_f,\Sigmab_f) \mbox{~avec~} \nonumber\\ 
\Sigmab_f &=& [\Phib^t <\Wb^t\Bf\Wb> \Phib+<\theta_f>\Db_f^t\Db_f]^{-1}, \nonumber\\ 
\mub_f    &=& \Sigmab_f \Phib^t  <\Wb>^t \Bf \gb,  
\label{eq45}
\eeqn
\beqn
q(\wb)    &=& \Nc(\mub_w,\Sigmab_w) \mbox{~avec~} \nonumber\\  
\Sigmab_w &=& [\Phib^t <\Fb^t\Bf\Fb> \Phib+\Ab]^{-1}, \nonumber\\ 
\mub_w    &=& \Sigmab_w \Phib^t  <\Fb>^t \Bf \gb 
\label{eq46}
\eeqn
\beqn
q(\theta_{e_i})        &=& \Gc(\alpha_{ei},\beta_{ei}) \mbox{~~avec~} \nonumber\\  
\alpha_{ei}            &=& \alpha_{e0} + M/2, \nonumber\\ 
\beta_{ei}             &=& \beta_{e0} + 1/2 <\epsilonb\epsilonb^t>_{ii}, 
\label{eq47}
\eeqn
\beqn
q(\theta_f)            &=& \Gc(\alpha_{f},\beta_{f}) \mbox{~~avec~} \nonumber\\ 
\alpha_{f}             &=& \alpha_{f0} + N/2,  \nonumber\\
\beta_{f}              &=& \beta_{f0} + 1/2 \, \trace{\Db_f^t\Db_f<\fb\fb^t>},\qquad \quad  
\label{eq48}
\eeqn
\beqn
q(\alpha_{wj})         &=& \Gc(a_{j},b_{j}) \mbox{~avec~} \nonumber\\ 
a_{j}            &=& \alpha_{w0} + 1/2, \nonumber \\
b_{j}            &=& \beta_{w0} + 1/2 <w_j^2>.
\label{eq49}
\eeqn
où 
\[
\barr{l}
\Ab=\diag{<\alpha_{wj}>, j=1,\cdots,N} \mbox{~et} \\
\Bb=\diag{<\beta_{ei}>, i=1,\cdots,M}.
\earr 
\]

On a ainsi l'expression des différentes composante de la loi séparable approchante. On peut en déduire facilement les moyennes de ces lois, car ces lois sont, soit des gaussiennes, soit des lois gamma. 
\beq
\barr{ll}
<\wb>=\mub_w, & <w_j^2>=[\mub_w]_j^2+[\Sigmab_w]_{jj}\\ 
<\fb>=\mub_f, & <\fb\fb^t>=\mub_f\mub_f^t + \Sigmab_f\\ 
<\theta_{ei}>=\alpha_{ei}/\beta_{ei} & <\theta_{f}>=\alpha_{f}/\beta_{f}, 
\earr
\label{eq50}
\eeq
et
\beq
\barr{ll}
<\epsilonb\epsilonb^t>&=\gb\gb^t-2\gb [<\Fb> \Phib <\wb>]^t \\ 
                      &+\Phib [\Fb <\wb><\wb>^t \Fb^t] \Phib^t
\earr
\label{eq51}
\eeq

Pour le calcul des termes $<\Wb^t\Wb>$ et $<\Fb^t\Fb>$ qui interviennent dans les expressions de $\Sigmab_f$, $\Sigmab_w$ et $[\Fb \wb\wb^t \Fb^t]$ on peut utiliser le fait que $\Fb$ et $\Wb$ sont des matrices bloc-Toeplitz avec des blocs Toeplitz (TBT), on peut les approcher par des matrices bloc-circulantes avec des blocs circulantes (CBC) et les inverser en utilisant la TFD. 
Notons aussi que $\mub_f$ et $\mub_w$ peuvent être obtenu par optimisation de 
\beq
J(\mub_f)=[\gb-\Phib\Wb \mub_f]^t\Bb[\gb-\Phib\Wb \mub_f]+<\theta_f>\|\Db_f\mub_f\|^2 \\ 
\label{eq52}
\eeq
et de 
\beq
J(\mub_w)=[\gb-\Phib\Fb \mub_w]^t\Bb[\gb-\Phib\Fb \mub_w]+\|\Ab\mub_w\|^2.  
\label{eq53}
\eeq

\section{Restauration avec modélisation Gauss-Markov-Potts}
Le cas d'une modélisation gaussienne reste assez restrictif pour la modélisation des images. 
Des modélisation par des champs de Markov composites (intensités-contours ou intensités-régions) 
sont mieux adaptées. Dans ce travail, nous examinons ce dernier. L'idée de base est 
de classer les pixels de l'images $\fb=\{f(\rb), \rb\in\Rc\}$ en $K$ classes par l'intermédiaire d'une variable discrète $z(\rb)\in\{1,\cdots,K\}$. L'image $\zb=\{z(\rb), \rb\in\Rc\}$ représente ainsi la segmentation de l'image $\fb$. 
Les pixels $\fb_k=\{f(\rb), \rb\in\Rc_k\}$ avec $\Rc_k=\{\rb ~:~ z(\rb)=k\}$ 
ont des propriétés communes, par exemple la même moyenne $\mu_k$ et la même variance $v_k$ (homogénéité au sens probabiliste). Ces pixels se trouvent dans un nombre fini de régions compactes et disjointes $\Rc_{kl}$ telles que~: 
$\cup_l \Rc_{kl}=\Rc_k$ et $\cup_k \Rc_{k}=\Rc$. 
On suppose aussi que $\fb_k$ et $\fb_l$ $\forall k\not=l$ sont indépendants.  

A chaque région est associée un contour. Si on représente les contours de l'images par 
une variable binaire $c(\rb)$, on a $c(\rb)=0$ à l'intérieure d'une région et $c(\rb)=1$ aux frontières de ces régions. On note aussi que $c(\rb)$ s'obtient à partir de $z(\rb)$ d'une manière déterministe (voir Fig~1). 

Avec cette introduction, nous pouvons définir
\beq
p(f(\rb)|z(\rb)=k,m_k,v_k)=\Nc(m_k,v_k) 
\label{eq54}
\eeq
ce qui suggère un modèle de mélange de gaussienne pour les pixels de l'image
\beq
p(f(\rb))=\sum_k a_k \Nc(m_k,v_k) \mbox{~avec~} a_k=P(z(\rb)=k)
\label{eq55}
\eeq
Une première modélisation simple est de supposer que les étiquettes $z(\rb)$  sont 
\aprio indépendants~:
\beq
p(\zb)=\prod_{\rb} p(z(\rb)).
\label{eq56}
\eeq
Nous appelons ce modèle, Mélange séparable de gaussiennes (MSG). 

Maintenant, pour prendre en compte la structure spatiale de ces pixels, nous devons introduire, d'une manière ou autre, une dépendance spatiale entre ces pixels. La modélisation markovienne est justement l'outil approprié. 

Cette dépendance spatiale peut être fait de trois manières. Soit utiliser un modèle markovien pour $z(\rb)$ et un modèle indépendant pour $f(\rb)|z(\rb)$, soit un modèle markovien pour $f(\rb)|z(\rb)$ et un modèle indépendant pour $z(\rb)$, soit un modèle markovien pour $f(\rb)|z(\rb)$ et un modèle markovien aussi pour $z(\rb)$. 
Nous avons examiné ces cas avec des modèles de Gauss-Markov pour $f(\rb)|z(\rb)$ et le modèle de Potts pour $z(\rb)$. Ce dernier peut s'écrire sous deux formes~:
\beq
p(z(\rb)|z(\rb'),\rb'\in\Vc(\rb))\propto 
\expf{\gamma \sum_{\rb'\in\Vc(\rb)}\delta(z(\rb)-z(\rb'))}
\label{eq57}
\eeq
et
\beq 
p(\zb)\propto 
\expf{\gamma \sum_{\rb\in\Rc}\sum_{\rb'\in\Vc(\rb)}\delta(z(\rb)-z(\rb'))}
\label{eq58}
\eeq

\bfig
\btabu{ccc}
\includegraphics[width=45mm,height=45mm]{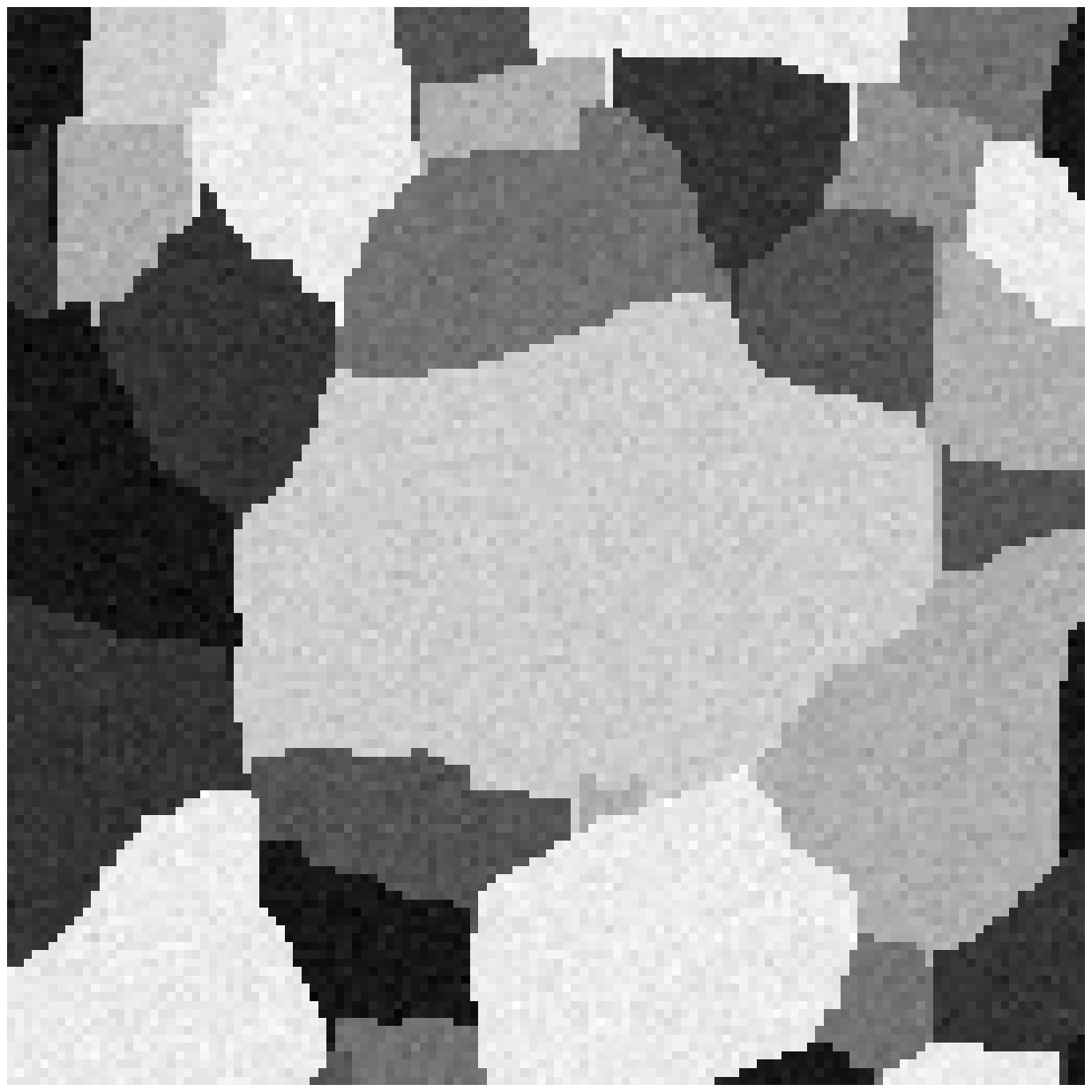}& 
\includegraphics[width=45mm,height=45mm]{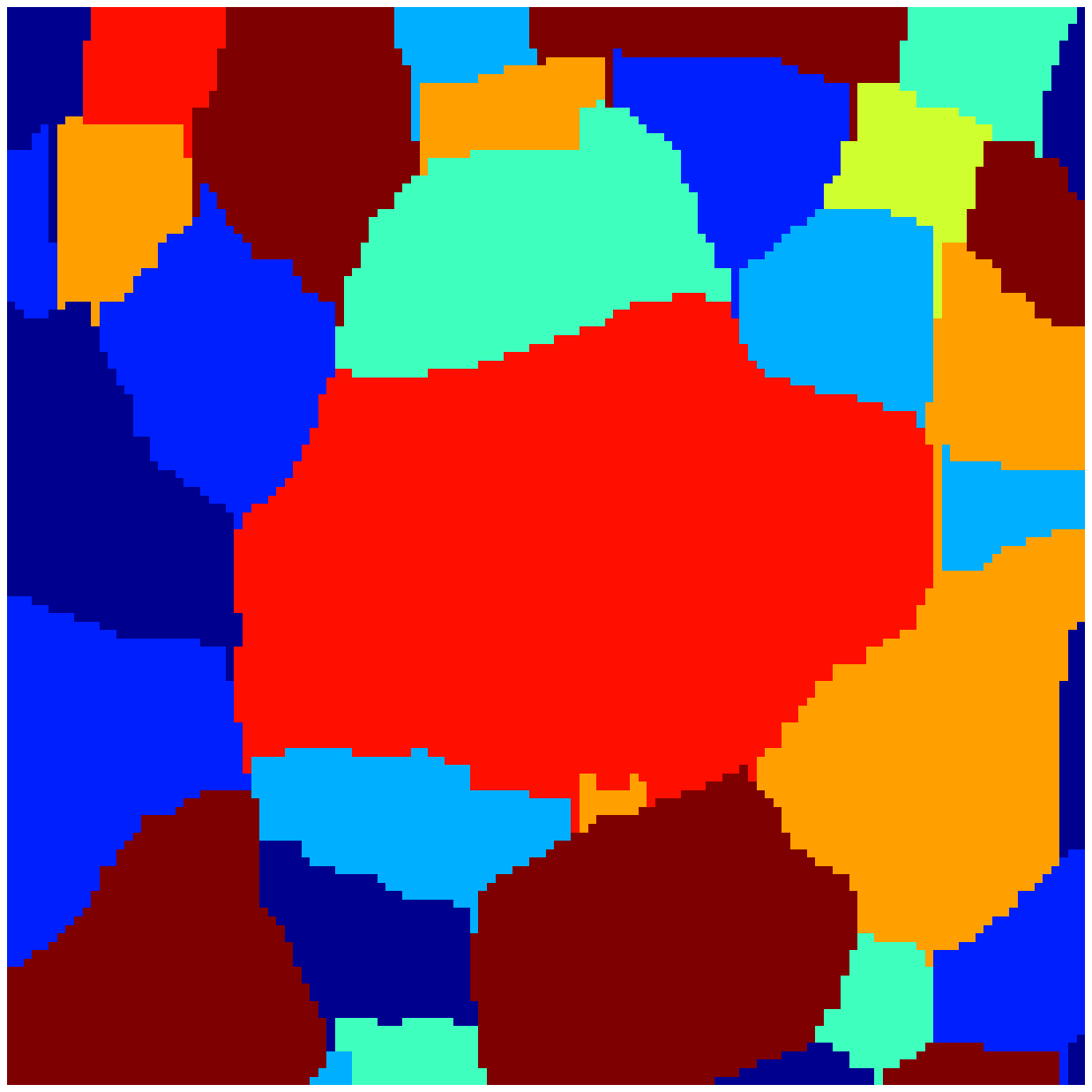}& 
\includegraphics[width=45mm,height=45mm]{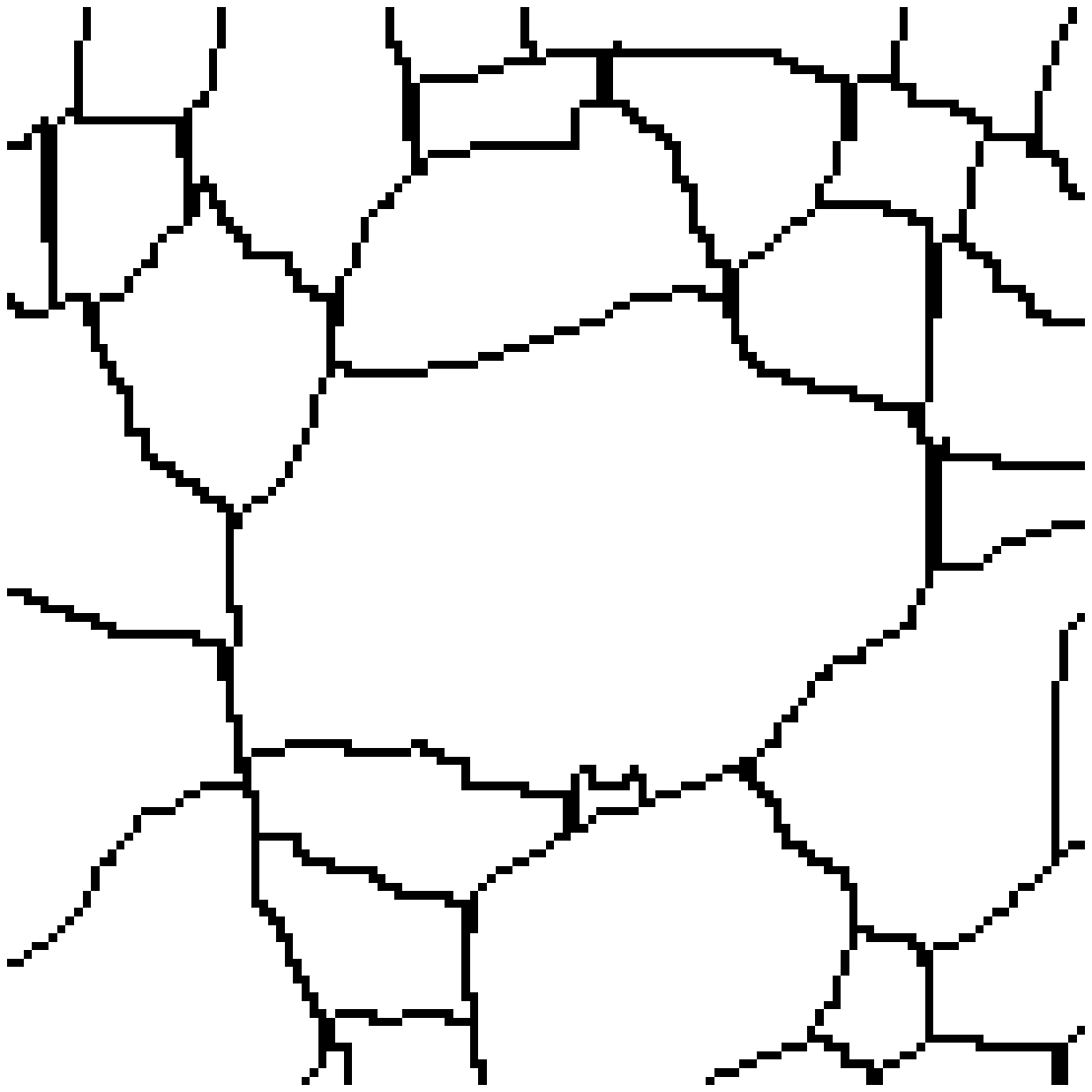}
\\ 
$f(\rb)$ & $z(\rb)$ & $c(\rb)$
\etabu
\caption[Modèle de mélange et champs de Markov caché]{Modèle de mélange et champs de Markov caché: image des intensités ou niveau de gris $f(\rb)$, image $z(\rb)$ de segmentation ou classification, image binaire $c(\rb)$ des contours.} 
\label{Fig1}
\efig

Ces différents cas peuvent alors se résumer par~:

\bigskip\noindent{\bf Modèle de mélange séparables de gaussiennes~(MSG):}\\ 
C'est le modèle le plus simple où aucune structure spatiale n'est pris en compte \aprio. Les relations qui donnent la loi \aprio conjointe de $p(\fb,\zb)=p(\fb|\zb) \; p(\zb)$ sont~:
\beq
\left\{\barr{lcl}
p(f(\rb)|z(\rb)=k)&=&\Nc(m_k,v_k), \forall \rb\in\Rc \\ 
p(\fb|\zb)        &=&\prod_{\rb\in\Rc} \Nc(m_z(\rb),v_z(\rb))  
\earr\right.
\label{eq59}
\eeq
avec $m_z(\rb)=m_k, \forall \rb\in\Rc_k$ et $v_z(\rb)=v_k, \forall \rb\in\Rc_k$,  
et 
\beq 
\left\{\barr{lcl}
p(z(\rb)=k)&=&\alpha_k, \; \forall \rb\in\Rc \\ 
p(\zb)     &=&\prod_{\rb} p(z(\rb)=k)=\prod_k \alpha_k^{n_k}
\earr\right.
\label{eq60}
\eeq
avec $n_k=\sum_{\rb\in\Rc} \delta(z(\rb)-k)$ le nombre de pixels dans la classe $k$ et $\sum_k n_k=n$ le nombre total des pixels de l'image. 
Les paramètres de ce modèle sont 
\[
\thetab_f=\{(\alpha_k,m_k,v_k), k=1,\cdots,K\}.
\] 
Rappelons aussi que 
$\sum_k \alpha_k=1$. Lors d'une estimation bayésienne non supervisée, il faut aussi attribuer des lois \aprio à ces paramètres. les lois conjuguées correspondantes sont des gaussiennes pour $m_k$, des Inverse Gammas pour $v_k$ et la loi Dirichlet pour 
$\alphab=[\alpha_k, k=1,\cdots,K]$~: 

\beq 
\left\{\barr{lcl}
p(m_k|m_0,v_0)         &=&\Nc(m_0,v_0), \; \forall k \\ 
p(v_k|\alpha_0,\beta_0)&=&\Ic\Gc(a_0,b_0), \; \forall k \\ 
p(\alphab|\alpha_0)    &=&\Dc(\alpha_0\oneb_K) 
\earr\right. 
\label{eq61}
\eeq
où $m_0, v_0, a_0, b_0$ et $\alpha_0$ sont fixés pour un problème donnée. On les choisis de telle sorte que ces lois soient les moins informatives (par exemple, $m_0=0$, $v_0=10$, $a_0=1$, $b_0=100$ et $\alpha_0=1/K$). 

\bigskip\noindent{\bf Modèle de mélange séparable de Gauss-Markov (MSGM):} \\ 
Ici, la structure spatiale est pris en compte au travers d'un modèle markovien sur $\fb$~: 
\beq
p(f(\rb)|z(\rb),f(\rb'),z(\rb'),\rb'\in\Vc(\rb))=\Nc(m_z(\rb),v_z(\rb))
\label{eq62}
\eeq
avec
\beq
\barr{lcl}
m_z(\rb)&=&\frac{1}{|\Vc(\rb)|}\sum_{\rb'\in\Vc(\rb)} m_{z(\rb')}^* 
\\ 
m_{z(\rb')}^*&=&\left\{\barr{lll}
\mu_{z(\rb')} & \mbox{si} & z(\rb')\not=z(\rb) \\ 
f(\rb')       & \mbox{si} & z(\rb')=z(\rb)  
\earr\right.
\earr
\label{eq63}
\eeq
Par contre le champs $\zb$ est supposé séparable comme dans le premier cas~(\ref{eq60}). 

On remarque que $\fb|\zb$ est un champ de Gauss-Markov non homogène car la moyenne $m_z(\rb)$ et la variance $v_z(\rb)$ varient en fonction de $\rb$. Tous les pixels $\fb_k=\{f(\rb), \rb\in\Rc_k\}$ se trouvant dans une région $k$ forment alors un vecteur gaussien de moyen $m_k\oneb_k$ et de matrice de covariance $\Sigmab_k$~: 
\beq
p(\fb_k)=\Nc(m_k\oneb_k,v_k\Sigmab_k)
\label{eq64}
\eeq
où $v_k\Sigmab_k$ est une matrice de covariance de dimension $n_k\times n_k$ et $\oneb_k=1$ est un vecteur de taille $n_k$ remplis de 1. Par ailleurs, on a aussi~: 
\beq
p(\fb|\zb)\propto \prod_{k} \Nc(m_k\oneb_k,v_k\Sigmab_k) 
\label{eq65}
\eeq
ce qui signifie que les pixels appartenant aux différentes régions sont indépendantes.

\bigskip\noindent{\bf Modèle de Gauss-Potts (MGP)~:} \\ 
Dans ce modèle, la structure spatiale est pris en compte au travers du modèle markovien de Potts~(\ref{eq57}) ou (\ref{eq58}). La loi de $p(\fb|\zb)$ est la même que dans le premier cas. En résumé, ici on a~: 
\beq
\left\{\barr{lcl}
p(\fb|\zb)     &=& \prod_{\rb\in\Rc} \Nc(m(\rb),v(\rb))\\ 
p(\zb)         &\propto& 
\expf{\gamma \sum_{\rb\in\Rc}\sum_{\rb'\in\Vc(\rb)}\delta(z(\rb)-z(\rb'))}
\earr\right.
\label{eq66}
\eeq
avec $m(\rb)=m_k, \forall \rb\in\Rc_k$ et $v(\rb)=v_k, \forall \rb\in\Rc_k$.   

Ici, donc, les pixels de l'images sont indépendantes conditionnellement à 
la connaissance des variables cachées, et donc, contrairement au cas précédent 
(\ref{eq64}), ici on a $p(\fb_k)=\Nc(m_k\oneb_k,v_k\Ib_k)$ où $v_k\Ib_k$ est 
une matrice de covariance identité de dimension $n_k\times n_k$, ce qui permet d'écrire~: 

\beq
p(\fb_k)=\Nc(m_k\oneb_k,v_k\Ib_k)
\propto\expf{\frac{-1}{2}\sum_{\rb\in\Rc_k}\frac{f(\rb)-m_k)^2}{ v_k}}
\label{eq67}
\eeq
et
\beqn
p(\fb|\zb)&\propto&\prod_{k} \Nc(m_k\oneb_k,\Ib_k) \nonumber \\ 
&\propto&\expf{\frac{-1}{2}\sum_k\sum_{\rb\in\Rc_k}\frac{(f(\rb)-m_k)^2}{v_k}}
\label{eq68}
\eeqn
ce qui signifie que, comme le cas précédent, les pixels appartenant aux différentes régions sont indépendantes. 

Les paramètres de ce modèle sont 
\(
\thetab_f=\{(\alpha_k,m_k,v_k), k=1,\cdots,K\}
\) 
et $\gamma$. Malheureusement, il n'y a pas de loi conjuguée pour ce paramètre et donc son estimation devient plus difficile. Dans ce travail, nous fixons ce paramètre \aprio. 

\bfig
\bcc
\btabu{cccc}
\includegraphics[width=30mm,height=30mm]{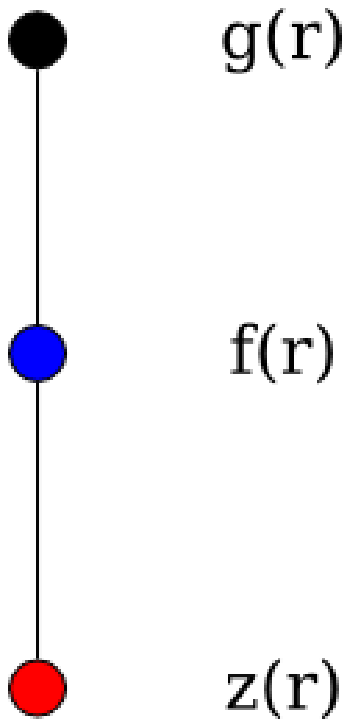}~~~~~~& 
\includegraphics[width=30mm,height=30mm]{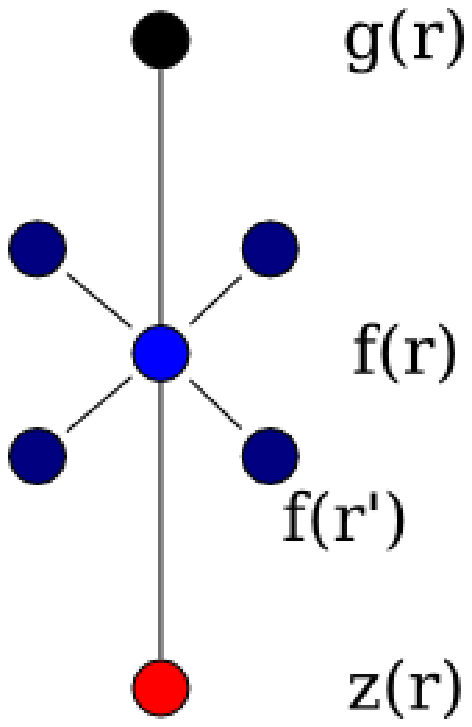}~~~~~~& 
\includegraphics[width=30mm,height=30mm]{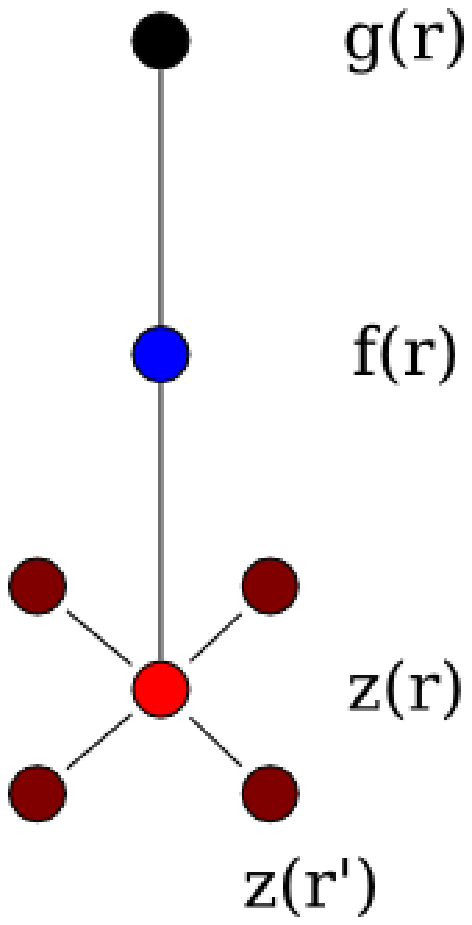}~~~~~~&
\includegraphics[width=30mm,height=30mm]{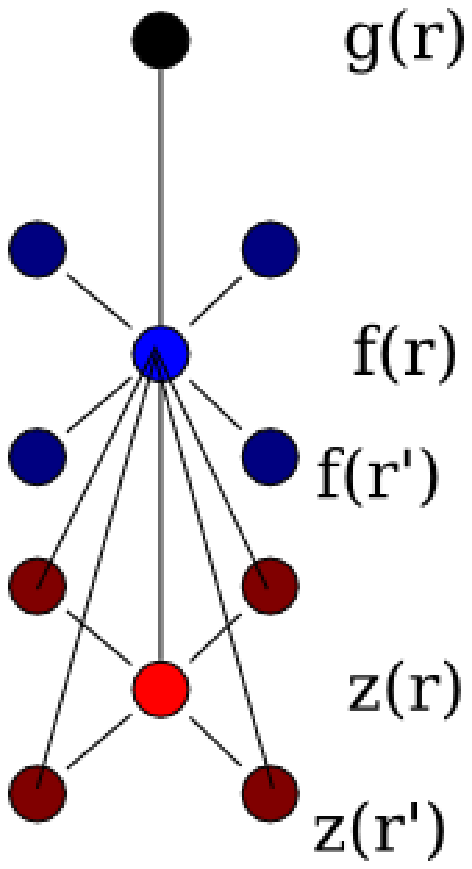}\\ 
MSG & MSGM & MGP & MGMP
\etabu
\ecc
\caption[Quatre modèles de mélange avec champs cachés]{Quatre modèles de mélange avec champs cachés. Dans le modèle MSG toutes les variables sont séparables en $\rb$. Dans MSGM, $f(\rb)$ dépends de $z(\rb)$ et de ses voisins $f(\rb')$. Dans MGP, $z(\rb)$ dépends de ses voisin $z(\rb')$. Dans MGMP, $z(\rb)$ dépends de ses voisin $z(\rb')$ et $f(\rb)$ dépends de $z(\rb)$, $z(\rb')$ et de ses voisins $f(\rb')$.} 
\label{Fig2}
\efig

\bigskip\noindent{\bf Modèle de Gauss-Markov-Potts (MGMP)~:} \\ 
Il s'agit là de la composition des deux derniers modèles. 
Ici, nous résumons ses relations d'une manière légèrement différente en utilisant~:
\beq
c_{\rb}(\rb')=1-\delta(z(\rb')-z(\rb))=\left\{\barr{lll}
1 & \mbox{si} & z(\rb')\not=z(\rb) \\ 
0 & \mbox{si} & z(\rb')=z(\rb)  
\earr\right.
\label{eq69}
\eeq
ce qui donne~:
\beq
p(f(\rb)|z(\rb),f(\rb'),z(\rb'),\rb'\in\Vc(\rb))=\Nc(m_z(\rb),v_z(\rb))
\label{eq70}
\eeq
avec
\beq
m_z(\rb)=\frac{1}{|\Vc(\rb)|}\sum_{\rb'\in\Vc(\rb)} \left[
(1-c_{\rb}(\rb')) \mu_{z(\rb')} + c_{\rb}(\rb') f(\rb')\right]
\label{eq71}
\eeq
et 
\beq
p(\zb)\propto
\expf{\gamma \sum_{\rb\in\Rc}\sum_{\rb'\in\Vc(\rb)}(1-c_{\rb}(\rb'))}
\label{eq72}
\eeq

\bigskip
En tous cas, quelque soit le modèle choisi parmi ces différents modèles, l'objectif est 
d'estimer $\fb$, $\zb$ et $\thetab$ en utilisant la loi \apost jointe~:
\beq
p(\fb,\zb,\thetab|\gb)=\frac{p(\gb|\fb,\thetab) \; p(\fb|\zb,\thetab) \; p(\zb)}{p(\gb|\thetab)}
\label{eq73}
\eeq
Bien que nous connaissons les expressions de tous les composants de numérateur de la fraction à la droite de cette relation, le calcul du dénominateur n'est que rarement possible d'une manière analytique.  
On cherche alors à approcher $p(\fb,\zb,\thetab|\gb)$ par un produit des lois plus simples à manipuler. Un premier choix est de l'approcher par une loi séparable  
$q(\fb,\zb,\thetab|\gb)=q_1(\fb)\;q_2(\zb)\; q_3(\thetab)$. 
Un deuxième choix est $q_1(\fb|\zb)\;q_2(\zb)\; q_3(\thetab)$ qui permet de garder des liens forts qui existent entre $\fb$ et $\zb$ et de ne relaxer que des liens faibles entre $\thetab$ et ces deux derniers. 

Le choix des familles appropriées pour $q_1(\fb|\zb)$ et $q_2(\zb)$ et $q_3(\thetab)$ pour chacun de ces modèles, qui est en lien avec les formes des lois \aprio dans chacun de ces cas, et les expressions de la mise à jour de ces différentes lois au cours des itérations nécessite beaucoup d'espace. 
Ici, nous allons juste fournir le principe et un résumé de ces relations.

Le choix de $q_3(\thetab)$, ou plus exactement, des familles des lois pour chacune des composentes de $\thetab$, qui sont la même pour tous les cas, sont des lois conjuguées de (\ref{eq61}). Par la suite, nous détaillons les choix de  $q(\fb|\zb)$ et $q(\zb)$ dans les différent cas. 

\medskip\noindent{\bf Cas 1 (MSG)~:} \\ 
Dans le cas du premier modèle 
on a 
\beq
\left\{\barr{lll}
p(\fb|\zb)&=&\prod_{\rb} p(f(\rb)|z(\rb)) \\ 
p(\zb)    &=&\prod_{\rb} p(z(\rb))  
\earr\right.
\label{eq74}
\eeq
où $p(f(\rb)|z(\rb)=k)=\Nc(m_k,v_k)$ et $p(z(\rb)=k)=\alpha_k$, 
ce qui naturellement nous conduit à choisir
\beq
\left\{\barr{lll}
q(\fb|\zb)&=&\prod_{\rb} q(f(\rb)|z(\rb)) \\ 
q(\zb)    &=&\prod_{\rb} q(z(\rb))  
\earr\right.
\label{eq75}
\eeq
avec $q(f(\rb)|z(\rb)=k)=\Nc(\tilde{m}_k,\tilde{v}_k)$ et $q(z(\rb)=k)=\tilde{\alpha}_k$. 

Au cours des itérations, $\tilde{m}_k$ $\tilde{v}_k$ et $\tilde{\alpha}_k$ seront mise à jour.

\medskip\noindent{\bf Cas 2 (MSGM)~:} \\ 
Dans le deuxième cas, le modèle de mélange de Gauss-Markov, 
notant que
\beq
p(\fb|\zb)=\prod_{\rb} \Nc(f(\rb)|m_z(\rb),v_z(\rb))
\label{eq76}
\eeq
où 
\beq
m_z(\rb)=\frac{1}{|\Vc(\rb)|}\sum_{\rb'\in\Vc(\rb)} m_{z(\rb')}^* 
\label{eq77}
\eeq
avec
\beq
m_{z(\rb')}^*=\delta(z(\rb')-z(\rb)) f(\rb') + (1-\delta(z(\rb')-z(\rb)) \mu_{z(\rb')}
\label{eq78}
\eeq
on trouve naturellement l'approximation suivante~:
\beq
m_{z(\rb')}^*=\delta(z(\rb')-z(\rb)) \tilde{f}(\rb') + (1-\delta(z(\rb')-z(\rb)) \tilde{m}_{z(\rb)}
\label{eq79}
\eeq
où $\tilde{f}(\rb')$ est l'espérance de $f(\rb')$ calculée à l'étape précédente. 
Il s'agit de l'approximation dite \emph{en champs moyens} pour le champ de Markov $\fb$~: 

\beq
\left\{\barr{@{}l@{}l@{}l@{}}
q(\fb|\zb)&=&\prod_{\rb} q(f(\rb)|z(\rb))
=\prod_{\rb} \Nc(f(\rb)|m_z(\rb),v_z(\rb)) \\ 
\mbox{avec} \\ 
m_z(\rb)&=&\frac{1}{|\Vc(\rb)|}\sum_{\rb'\in\Vc(\rb)} \\ 
&& \delta(z(\rb)-z(\rb')) f(\rb) + (1-\delta(z(\rb)-z(\rb'))) \mu_{z(\rb)}
\earr\right.
\label{eq80}
\eeq

\medskip\noindent{\bf Cas 3 (MGP)~:} \\ 
Dans le troisième cas, le modèle de Gauss-Potts 
on a 
\beq
\left\{\barr{lll}
p(\fb|\zb)&=&\prod_{\rb} p(f(\rb)|z(\rb)) \\ 
p(\zb)    &=&\prod_{\rb} p(z(\rb)|z(\rb'), \rb'\in\Vc(\rb))  
\earr\right.
\label{eq81}
\eeq
ce qui naturellement nous conduit à choisir
\beq
\left\{\barr{lll}
q(\fb|\zb)&=&\prod_{\rb} q(f(\rb)|z(\rb)) \\ 
q(\zb)    &=&\prod_{\rb} q(z(\rb)|\bar{z}(\rb'), \rb'\in\Vc(\rb))  
\earr\right.
\label{eq82}
\eeq
où nous avons choisi une approximation en champs moyens pour le champs de Potts. 
Cette approximation supplémentaire qui consiste à remplacer les valeurs de $z(\rb')$ par leurs moyenne $\bar{z}(\rb')$ est une approximation courante pour un modèle de Potts. 

\medskip\noindent{\bf Cas 4 (MGMP)~:} \\ 
Dans le quatrième cas,le Modèle de Gauss-Markov-Potts 
on a 
\beq
\left\{\barr{@{}l@{}l@{}l@{}}
q(\fb|\zb)&=&\prod_{\rb} q(f(\rb)|z(\rb))
=\prod_{\rb} \Nc(f(\rb)|m_z(\rb),v_z(\rb)) \\ 
\mbox{avec} \\ 
m_z(\rb)&=&\frac{1}{|\Vc(\rb)|}\sum_{\rb'\in\Vc(\rb)} \\ 
&& \delta(z(\rb)-z(\rb') f(\rb) + (1-\delta(z(\rb)-z(\rb')) \mu_{z(\rb)}
\\ 
q(\zb)    &=&\prod_{\rb} q(z(\rb)|\bar{z}(\rb'), \rb'\in\Vc(\rb))  
\earr\right.
\label{eq83}
\eeq

\bigskip 
Dans tous les cas, puisque les lois \aprio pour les différent composants de $\thetab=\{\{\theta_{e_i}\},\{m_{k}\},\{v_{k}\},\{\beta_{k}\}\}$ 
sont des lois conjuguées et séparables, on choisi
\beq
q(\thetab)=\prod_i q(\theta_i)
\label{eq84}
\eeq

Il reste un paramètre que nous garderons fixe. C'est le paramètre du modèle de Potts. En effet, n'ayant pas une expression analytique pour la fonction de répartition du modèle de Potts, il n'existe pas une loi conjuguée pour ce paramètre. Dans un premier temps donc, nous gardons ce paramètre comme un paramètre de réglage non supervisé. 

Les deux tableaux qui suivent résument les lois \aprio et les choix des lois séparables pour les quatre cas proposés. 

\begin{table}
\hspace*{-1.9cm}
\btabu{|c|c|c|c|c|} \hline 
Modèle & MSG & MSGM & MGP & MGMP \\ \hline 
$p(\fb|\zb)$ & $\disp{\prod_{\rb}\Nc(m_z(\rb),v_z(\rb))}$ & Gauss-Markov  &  $\disp{\prod_{\rb}\Nc(m_z(\rb),v_z(\rb))}$   &   Gauss-Markov    
\\ \hline 
$p(\zb)$ & $\disp{\prod_{\rb}\alpha_z(\rb)=\prod_k\alpha_k^{n_k}}$ & $\disp{\prod_{\rb}\alpha_z(\rb)=\prod_k\alpha_k^{n_k}}$  &  Potts   &   Potts    
\\ \hline
$p(m_k)$ & $\Nc(m_0,v_0)$ & $\Nc(m_0,v_0)$  &  $\Nc(m_0,v_0)$   &   $\Nc(m_0,v_0)$   
\\ \hline
$p(v_k)$ & $\Ic\Gc(a_0,b_0)$ & $\Ic\Gc(a_0,b_0)$  &  $\Ic\Gc(a_0,b_0)$   &   $\Ic\Gc(a_0,b_0)$   
\\ \hline
$p(\alphab)$ & $\Dc(\alpha_0,\cdots,\alpha_0)$ & $\Dc(\alpha_0,\cdots,\alpha_0)$ &  $\Dc(\alpha_0,\cdots,\alpha_0)$ & $\Dc(\alpha_0,\cdots,\alpha_0)$   
\\ \hline
$p(\theta_e)$ & $\Gc(a_{\epsilon_0},b_{\epsilon_0})$ & $\Gc(a_{\epsilon_0},b_{\epsilon_0})$  &  $\Gc(a_{\epsilon_0},b_{\epsilon_0})$   &   $\Gc(a_{\epsilon_0},b_{\epsilon_0})$   
\\ \hline\hline
$q(\fb|\zb)$ & $\disp{\prod_{\rb}\Nc(\hat{\mu}_z(\rb),\hat{v}_z(\rb))}$ & $\disp{\prod_{\rb}\Nc(\hat{\mu}_z(\rb),\hat{v}_z(\rb))}$  &  $\disp{\prod_{\rb}\Nc(m_z(\rb),v_z(\rb))}$   &   $\disp{\prod_{\rb}\Nc(\hat{\mu}_z(\rb),\hat{v}_z(\rb))}$    
\\ \hline 
$q(\zb)$ & $\disp{\prod_{\rb}\hat{\alpha}_z(\rb)=\prod_k\hat{\alpha}_k^{n_k}}$ & $\disp{\prod_{\rb}\hat{\alpha}_z(\rb)=\prod_k\hat{\alpha}_k^{n_k}}$  &  
$\disp{\prod_{\rb} q(z(\rb)|\bar{z}(\rb'), \rb'\in\Vc(\rb))}$   &   
$\disp{\prod_{\rb} q(z(\rb)|\bar{z}(\rb'), \rb'\in\Vc(\rb))}$    
\\ \hline
$q(m_k)$ & $\Nc(\hat{m}_k,\hat{v}_k)$ & $\Nc(\hat{m}_k,\hat{v}_k)$  &  $\Nc(\hat{m}_k,\hat{v}_k)$   &   $\Nc(\hat{m}_k,\hat{v}_k)$   
\\ \hline
$q(v_k)$ & $\Ic\Gc(\hat{a}_k,\hat{b}_k)$ & $\Ic\Gc(\hat{a}_k,\hat{b}_k)$  &  $\Ic\Gc(\hat{a}_k,\hat{b}_k)$   &   $\Ic\Gc(\hat{a}_k,\hat{b}_k)$   
\\ \hline
$q(\alphab)$ & $\Dc(\hat{\alpha}_1,\cdots,\hat{\alpha}_K)$ & $\Dc(\hat{\alpha}_1,\cdots,\hat{\alpha}_K)$ &  $\Dc(\hat{\alpha}_1,\cdots,\hat{\alpha}_K)$ & $\Dc(\hat{\alpha}_1,\cdots,\hat{\alpha}_K)$   
\\ \hline
$q(\theta_e)$ & $\Gc(\hat{a}_{\epsilon_0},\hat{b}_{\epsilon_0})$ & $\Gc(\hat{a}_{\epsilon_0},\hat{b}_{\epsilon_0})$  &  $\Gc(\hat{a}_{\epsilon_0},\hat{b}_{\epsilon_0})$   &   $\Gc(\hat{a}_{\epsilon_0},\hat{b}_{\epsilon_0})$   
\\ \hline
\etabu
\caption{Les lois \aprio  et les lois séparables approchantes pour les quatre modèles proposés.}
\end{table}

Ici, nous ne développons pas plus les expressions de la mise à jour de ces différentes loi au cours des itérations, mais ce qu'il faut savoir est que toutes ces lois étant de formes paramétriques connues (gaussienne, gamma ou inverse gamma, Wishart ou inverse Wishart et Dirichlet), nous avons des expressions analytiques pour les espérances de ces lois. Les détails de ces relations accompagnés des résultats de simulation seront communiqués dans un autre papier dans un future proche.

\section{Résultats et discussions}
Nous avons utilisé cette approche dans plusieur domaines de problèmes inverses~: 
i) restauration d'image \cite{Ayasso2007a,Ayasso2008}, 
ii) reconstruction d'image en tomographie X \cite{Ayasso2008c,Ayasso2009a}, et  
iii) séparation de sources et segmentation des images hyperspectrales \cite{Bali2008}. 
Nous sommes aussi en train de mettre en oeuvre ces algorithmes pour le cas 3D de la tomographie X ainsi qu'au cas de l'imagerie microondes. 


Dans tous ces applications, nous avons implémenté à la fois des algorithmes MCMC (échantillonneurs de Gibbs) et l'approche variationnelle bayésienne (VB) développée dans cet article. Nous avons ainsi pu comparé les résultats obtenus par ces deux méthodes. 
D'une manière générale, les principales conclusions de ces expérimentations et comparaisons sont~: 
i) la qualité des résultats obtenus sont pratiquement similaires, et 
ii) le principal avantage de l'approche VB est dans le gain en temps de calcul qui dépend bien sure du contexte et de l'applications.  

Un autre avantage important que l'on peut mentionner est l'existance d'un critère (l'énergie libre, équation~\ref{eq10}) que l'on peut utiliser~: 
i) comme critère d'arrêt pour l'algorithme et 
ii) comme un critère de choix de modèle. 
En effet, comme nous l'avons vu, grâce à la relation~\ref{eq09}, minimiser $\mbox{KL}(q:p)$ est équivalent à maximiser $\Fc(q)$ et sa valeur optimale est un bon indicateur pour $\ln p(\gb|\Mc)$. 
Ainsi, les valeurs relatives de $\Fc(q)$ au cours des itérations de l'algorithme peuvent être utilisées comme un critère d'arrêt et sa valeur optimale atteint pour un modèle $\Mc_1$ peut être comparée à sa valeur optimale atteint pour un autre modèle $\Mc_2$ comme un critère de préférence entre les deux modèles. 
 
\section{Conclusion}
L'approche variationnelle de l'approximation d'une loi par des lois séparables  est appliquée au 
cas de l'estimation non supervisée des inconnues et des hyperparamètres dans des problèmes 
inverses de restauration d'image (déconvolution simple ou aveugle) avec des modélisations \aprio gaussiennes, 
gaussiennes généralisées, mélange de gaussiennes indépendantes ou mélange de Gauss-Markov avec champs cachés des étiquettes indépendantes ou markovien (Potts).  

Dans ce papier, nous avons décrit d'abord le principe de l'approximation d'une loi et ensuite proposé des approximations appropriées pour des lois \apost conjointes que l'on trouve dans le cas des problèmes inverses en général et en restauration d'images en particulier. Cependant, la mise en oeuvre effective de ces méthodes est en cours et les résultats de simulation et évaluation des performances de ces méthodes seront communiqués dans un future proche. 

Les détails de ces relations accompagnés des résultats de simulation qui sont disponible sous forme d'un rapport \cite{Ayasso2007a,Ayasso2008} seront communiqués dans un autre papier en préparation dans un futur proche. 

\section{Remerciment}
L'author souhaite remercier Jean-François Berchet et Guy Demoment pour les discussions fructueuses et la relecture de la première version de ce papier, H. Ayasso et S. Fekih pour l'implémentation de ces algorithmes en restauration d'image et en tomographie X, N. Bali  pour l'implémentation de ces algorithmes en séparation de sources et en imagerie hyperspectrale. 

{\small 
\bibliographystyle{ieeetr}
\def\bibdira{/home/djafari/Tex/Inputs/bib/commun/}
\def\bibdirb{/home/djafari/Tex/Inputs/bib/}
\def\bibdirc{/home/djafari/Tex/Inputs/bib/amd/}
\def\bibdir{/home/seismic/TeX/biblio/}
\def\UP#1{\uppercase{#1}}
\bibliography{\bibdira revuedef,\bibdira bibdef,\bibdira baseAJ,\bibdira baseKZ,\bibdirc Ali_Mohammad-Djafari,varBayes,\bibdirb aistats2005,\bibdirb references}
}

\end{document}

%% file: macros1.tex
\def\bdoc{\begin{document}}             \def\edoc{\end{document}}
\def\babs{\begin{abstract}}             \def\eabs{\end{abstract}}
\def\bcc{\begin{center}}                \def\ecc{\end{center}}
\def\ben{\begin{enumerate}}             \def\een{\end{enumerate}}
\def\bit{\begin{itemize}}               \def\eit{\end{itemize}}
\def\bpic{\begin{picture}}              \def\epic{\end{picture}}
\def\beq{\begin{equation}}              \def\eeq{\end{equation}}
\def\barr{\begin{array}}                \def\earr{\end{array}}
\def\btab{\begin{tabular}}              \def\etab{\end{tabular}}
\def\btabu{\begin{tabular}}             \def\etabu{\end{tabular}}
\def\bcc{\begin{center}}                \def\ecc{\end{center}}
\def\beqn{\begin{eqnarray}}             \def\eeqn{\end{eqnarray}}
\def\beqnx{\begin{eqnarray*}}           \def\eeqnx{\end{eqnarray*}}
\def\bfig{\begin{figure}}               \def\efig{\end{figure}}
\def\bver{\begin{verb}}						 \def\ever{\end{verb}}

\def\d#1{\,\mbox{d} #1}
\def\disp#1{\displaystyle{#1}}

\def\iii{\int_{-\infty}^{+\infty}}
\def\izi{\int_{0}^{\infty}}
\def\izpi{\int_{0}^{\pi}}
\def\izdpi{\int_{0}^{2\pi}}
\def\intd{\int\kern-.8em\int}
\def\intt{\int\kern-.8em\int\kern-.8em\int}
\def\intg{\int\kern-1.1em\int}

\def\xh{\widehat{x}}
\def\thetah{\widehat{\theta}}
\def\betah{\widehat{\beta}}

\def\xbh{\widehat{\xb}}
\def\thetabh{\widehat{\thetab}}
\def\betabh{\widehat{\betab}}

\def\xbhk{\widehat{\xb}^{k}}
\def\thetahk{\widehat{\theta}^{k}}
\def\betahk{\widehat{\beta}^{k}}

\def\xbhkp{\widehat{\xb}^{k+1}}
\def\thetahkp{\widehat{\theta}^{k+1}}
\def\betahkp{\widehat{\beta}^{k+1}}

\def\thetabhk{\widehat{\thetab}^{k}}
\def\betabhk{\widehat{\betab}^{k}}

\def\thetabhkp{\widehat{\thetab}^{k+1}}
\def\betabhkp{\widehat{\betab}^{k+1}}

\def\thetamin{\theta_{\mbox{\tiny min}}}
\def\thetamax{\theta_{\mbox{\tiny max}}}
\def\betamin{\beta_{\mbox{\tiny min}}}
\def\betamax{\beta_{\mbox{\tiny max}}}

\def\ra{\rightarrow}
\def\la{\leftarrow}
\def\da{\downarrow}
\def\ua{\uparrow}

\def\Ra{\Rightarrow}
\def\La{\Leftarrow}
\def\Da{\Downarrow}
\def\Ua{\Uparrow}

\def\lra{\longrightarrow}
\def\lla{\longleftarrow}
\def\Lra{\Longrightarrow}
\def\Lla{\longleftarrow}

\def\lrarr{\leftrightarrow}
\def\Lrarr{\Leftrightarrow}
\def\udarr{\updownarrow}
\def\Uparr{\Updownarrow}

\def\expf#1{\exp\left[ {#1} \right]}

\def\argmax#1#2{\arg\max_{#1}\left\{ #2 \right\}}
\def\argmin#1#2{\arg\min_{#1}\left\{ #2 \right\}}

\def\Prob#1{\mbox{Pr}\left\{#1\right\}}
\def\var#1{\mbox{Var}\left\{#1\right\}}
\def\cov#1{\mbox{Cov}\left\{#1\right\}}
\def\corr#1{\mbox{Corr}\left\{#1\right\}}
\def\trace#1{\mbox{Tr}\left\{#1\right\}}
\def\rang#1{\mbox{rang}\left\{#1\right\}}
\def\det#1{\mbox{d\'et}\left\{#1\right\}}

\def\gbu{\underline{\gb}}
\def\fbu{\underline{\fb}}
\def\zbu{\underline{\zb}}
\def\epsilonbu{\underline{\epsilonb}}
\def\thetabu{\underline{\thetab}}

\def\sigmae{{\sigma_{\epsilon}}}
\def\Sigmae{{\Sigmab_{\epsilonb}}}
\def\Sigmaf{{\Sigmab_{\fb}}}
\def\Sigmag{{\Sigmab_{\gb}}}

\def\fbuh{\widehat{\fbu}}
\def\zbuh{\widehat{\fbu}}
\def\thetabuh{\widehat{\thetabu}}

\def\fbh{\widehat{\fb}}
\def\Pbh{\widehat{\Pb}}
\def\Sigmabh{\widehat{\Sigmab}}
\def\Abh{\widehat{\Ab}}
\def\Bbh{\widehat{\Bb}}
\def\thetabh{\widehat{\thetab}}
\def\Rbe{\Rb_{\epsilon}}
\def\Rbeh{\widehat{\Rbe}}

\def\Rjk{{R_j}_k}
\def\fbik{{\fb_i}_k}
\def\fbjk{{\fb_j}_k}

\def\mik{{m_i}_k}
\def\sigmaik{{\sigma_i^2}_k}
\def\Sigmaik{{\Sigmab_i}_k}
\def\alphaik{{\alpha_i}_k}
\def\betaik{{\beta_i}_k}

\def\muik{{\mu_i}_k}
\def\vik{{v_i}_k}

\def\mjk{{m_j}_k}
\def\sigmajk{{\sigma_j^2}_k}
\def\Sigmajk{{\Sigmab_j}_k}
\def\alphajk{{\alpha_j}_k}
\def\betajk{{\beta_j}_k}

\def\mujk{{\mu_j}_k}
\def\vjk{{v_j}_k}
\def\njk{{n_j}_k}

\def\alphae{\alpha_{\epsilon}}
\def\betae{\beta_{\epsilon}}
\def\alphaez{\alpha_{\epsilon_0}}
\def\betaez{\beta_{\epsilon_0}}
\def\alphaei{\alpha_{\epsilon_i}}
\def\betaei{\beta_{\epsilon_i}}
\def\alphaeiz{\alpha_{\epsilon_{i_0}}}
\def\betaeiz{\beta_{\epsilon_{i-0}}}

\def\mbz{{{\mb}_z}}
\def\Sigmabz{{{\Sigmab}_z}}
\def\diag#1{\mbox{diag}\left[#1\right]}

\def\sigmah{\widehat{\sigma}}
\def\fh{\widehat{f}}
\def\sigmaz{\sigma_z}

\def\blue#1{{\color{blue}#1}}
\def\red#1{{\color{red}#1}}
\def\green#1{{\color{green}#1}}

\def\sumi{\sum_{i=1}^{M} }
\def\sumj{\sum_{j=1}^{N} }
\def\lambdah{\wh{\lambda}}
\def\lambdabh{\wh{\lambdab}}

\def\det#1{\hbox{det}\left(#1\right)}
\def\defined{\,\shortstack{$\triangle$\\ =}\,}

\def\zmat{\left[\matrix{0}\right]}
\def\det#1{\hbox{det}(#1)}
\def\th{\widehat{t}}
\def\xh{\widehat{x}}
\def\lambdah{\widehat{\lambda}}
\def\muh{\widehat{\mu}}
\def\sigmah{\widehat{\sigma}}
\def\rhoh{\widehat{\rho}}
\def\betah{\widehat{\beta}}
\def\alphah{\widehat{\alpha}}
\def\tbh{\widehat{\tb}}
\def\mubh{\widehat{\mub}}
\def\sigmabh{\widehat{\sigmab}}
\def\rhobh{\widehat{\rhob}}
\def\oneb{\mbox{\bf 1}}

\def\bm#1{\mbox{\boldmath $#1$}}
\def\zerob{{\bf 0}}
\def\oneb{{\bf 1}}
\def\intg{\int\kern-1.1em\int}
\def\expf#1{\exp\left[ {#1} \right]}

\def\gbu{\underline{\gb}}
\def\fbu{\underline{\fb}}
\def\zbu{\underline{\zb}}
\def\epsilonbu{\underline{\epsilonb}}
\def\uthetab{\underline{\thetab}}

\def\sigmae{{\sigma_{\epsilon}}}
\def\Sigmae{{\Sigma_{\epsilonb}}}
\def\Sigmabe{{\Sigmab_{\epsilonb}}}
\def\Sigmaf{{\Sigmab_{\fb}}}
\def\Sigmag{{\Sigmab_{\gb}}}

\def\fbuh{\widehat{\fbu}}
\def\zbuh{\widehat{\fbu}}
\def\uthetabh{\widehat{\uthetab}}

\def\fbh{\widehat{\fb}}
\def\Sigmabh{\widehat{\Sigmab}}
\def\Abh{\widehat{\Ab}}
\def\thetabh{\widehat{\thetab}}

\def\Rjk{{R_j}_k}
\def\fbik{{\fb_i}_k}
\def\fbjk{{\fb_j}_k}

\def\mik{{m_i}_k}
\def\sigmaik{{\sigma_i^2}_k}
\def\Sigmaik{{\Sigmab_i}_k}
\def\alphaik{{\alpha_i}_k}
\def\betaik{{\beta_i}_k}

\def\muik{{\mu_i}_k}
\def\vik{{v_i}_k}

\def\mjk{{m_j}_k}
\def\sigmajk{{\sigma_j^2}_k}
\def\Sigmajk{{\Sigmab_j}_k}
\def\alphajk{{\alpha_j}_k}
\def\betajk{{\beta_j}_k}

\def\mujk{{\mu_j}_k}
\def\vjk{{v_j}_k}
\def\njk{{n_j}_k}

\def\mbz{{{\mb}_z}}
\def\Sigmabz{{{\Sigmab}_z}}

\def\vect{\mbox{vect}}
\def\lra{\longrightarrow}

\def\gir{g_i(\rb)}
\def\fjr{f_j(\rb)}
\def\zjr{z_j(\rb)}
\def\epsilonir{\epsilon_i(\rb)}
\def\gbr{\gb(\rb)}
\def\fbr{\fb(\rb)}
\def\zbr{\zb(\rb)}
\def\epsilonbr{\epsilonb(\rb)}
\def\fbhr{\fbh(\rb)}
\def\Sigmabhr{\Sigmabh(\rb)}

\def\mbzr{\mb_{z(\rb)}}
\def\Sigmabzr{\Sigmab_{z(\rb)}}
\def\Sigmagbzr{{\Sigmab_g}_{z(\rb)}}

\def\mjzjr#1{{m_{#1}}_{z_{#1}(\rb)}}
\def\sigmajzjr#1{{\sigma_{#1}}_{z_{#1}(\rb)}}

\def\Beta{B}
\def\Betab{\bm{\Beta}}
\def\Betabe{\bm{\Beta}_{\epsilonb}}

\def\fh{\widehat{f}}
\def\sigmah{\widehat{\sigma}}
\def\sigmaei{\sigma_{\epsilon_i}^2}

\def\thetah{\widehat{\theta}}
\def\sigmah{\widehat{\sigma}}

\def\Ftxm{\widetilde{F}_{X}(x_-)}
\def\Ftxp{\widetilde{F}_{X}(x_+)}

\def\NU{{\cal V}}

\def\tFx{\widetilde{F}_{X}}

\def\Fx{F_{X}(x)}
\def\fx{f_{X}(x)}

\def\Fxv{F_{X}(x)}
\def\fxv{f_{X}(x)}

\def\Fxi{{F}_{X_i}(x_i)}
\def\fxi{{f}_{X_i}(x_i)}

\def\Gnu{G_{\NU}(\nu)}
\def\gnu{g_{\NU}(\nu)}

\def\Fnu{F_{\NU}(\nu)}
\def\fnu{f_{\NU}(\nu)}

\def\Finnu#1{F^{-1}_{\NU}(#1)}

\def\Gnua#1{G_{\NU}(#1)}
\def\Gnub#1{G_{\NU}^{-1}(#1)}
\def\Gnuh{G_{\NU}^{-1}(1/2)}

\def\Ftx{\widetilde{F}_{X}(x)}
\def\ftx{\widetilde{f}_{X}(x)}

\def\Ftxi{\widetilde{F}_{X_i}(x_i)}
\def\ftxi{\widetilde{f}_{X_i}(x_i)}

\def\Ftxv{\widetilde{F}_{X}(x)}
\def\FtX#1{\widetilde{F}_{X}(#1)}
\def\ftxv{\widetilde{f}_{X}(x)}

\def\fxnu{f_{X|\NU}(x|\nu)}
\def\fxNU{f_{X|\NU}(x|\NU)}
\def\FxNU{F_{X|\NU}(x|\NU)}

\def\fxNUtheta{f_{X|\NU,\theta}(x|\NU,\theta)}
\def\FxNUtheta{F_{X|\NU,\theta}(x|\NU,\theta)}

\def\Fxnu{F_{X|\NU}(x|\nu)}
\def\Fxnui#1{F_{X|\NU}(x|\nu_#1)}

\def\FxN#1{F_{X|\NU}(#1)}
\def\FxNU{F_{X|\NU}(x|\NU)}
\def\Fxnun#1{F_{X|\NU}(x|#1)}
\def\Fxinun#1{F_{X_i|\NU}(x_i|#1)}
\def\fxnu{f_{X|\NU}(x|\nu)}
\def\fxNU{f_{X|\NU}(x|\NU)}

\def\Fxvnu{F_{X|\NU}(x|\nu)}
\def\Fxvnun#1{F_{X|\NU}(x|#1)}
\def\FxvNU{F_{X|\NU}(x|\NU)}

\def\FXvnu{F_{X|\NU}}

\def\fxvnu{f_{X|\NU}(x|\nu)}
\def\fxvNU{f_{X|\NU}(x|\NU)}

\def\FXvNU#1{{F}_{X|\NU}(#1|\NU)}
\def\FXvn#1{{F}_{X|\NU}(x|#1)}

\def\Fxtheta{F_{X|\theta}(x|\theta)}
\def\fxtheta{f_{X|\theta}(x|\theta)}

\def\Fxvtheta{F_{X|\theta}(x|\theta)}
\def\fxvtheta{f_{X|\theta}(x|\theta)}

\def\Ftxtheta{\widetilde{F}_{X|\theta}(x|\theta)}
\def\ftxtheta{\widetilde{f}_{X|\theta}(x|\theta)}

\def\Ftxvtheta{\widetilde{F}_{X|\theta}(x|\theta)}
\def\ftxvtheta{\widetilde{f}_{X|\theta}(x|\theta)}

\def\Fxnutheta{F_{X|\NU,\theta}(x|\nu,\theta)}
\def\fxnutheta{f_{X|\NU,\theta}(x|\nu,\theta)}

\def\ftheta{f_{\Theta}(\theta)}
\def\fthetaxnuz{f_{\Theta|X,\nu_0}(\theta|x,\nu_0)}
\def\fthetax{f_{\Theta|X}(\theta|x)}

\def\Fxvnutheta{F_{X|\NU,\theta}(x|\nu,\theta)}
\def\FxvNUtheta{F_{X|\NU,\theta}(x|\NU,\theta)}
\def\fxvnutheta{f_{X|\NU,\theta}(x|\nu,\theta)}
\def\Ftxvnutheta{\widetilde{F}_{X|\theta}(x|\theta)}
\def\ftxvnutheta{\widetilde{f}_{X|\theta}(x|\theta)}

\def\FxvNUm{F_{X|\NU}(x_-|\NU)}
\def\FxvNUp{F_{X|\NU}(x_+|\NU)}

\def\Fxinu{F_{X_i|\NU}(x_i|\nu)}
\def\fxinu{f_{X_i|\NU}(x_i|\nu)}

\def\Ftxi{\widetilde{F}_{X_i}(x_i)}
\def\ftxi{\widetilde{f}_{X_i}(x_i)}

\def\fxitheta{f_{X_i|\theta}(x_i|\theta)}
\def\ftxitheta{\widetilde{f}_{X_i|\theta}(x_i|\theta)}

\def\thetah{\widehat{\theta}}
\def\thetat{\widetilde{\theta}}

\def\lnuxv{l_{\nu|X}(\nu|x)}

\def\Lnuxva#1{L_{\nu|X}(#1 | x)}
\def\lnuxva#1{l_{\nu|X}(#1 | x)}

\def\Prob#1{\mbox{P}\left(#1\right)}

\def\lthetaxvnuz{l(\theta)}

\def\lthetaxvnuz{l_{\theta|x,\nu_0}(\theta | x,\nu_0)}
\def\fxnuztheta{f_{X|\nu_0,\theta}(x | \nu_0,\theta)}
\def\fxinuztheta{f_{X|\nu_0,\theta}(x_i | \nu_0,\theta)}
\def\fxvtheta{f_{X|\theta}(x | \theta)}
\def\lthetaxv{l_{\theta|X}(\theta | x)}
\def\ltthetaxv{\widetilde{l}_{\theta|X}(\theta | x)}

\def\ltheta{l(\theta )}
\def\lttheta{\widetilde{l}(\theta)}
\def\Lnu{L(\nu )}
\def\Lnui#1{L(\nu_#1)}
\def\Linu{L_i(\nu )}
\def\Linu{L_i(\nu )}
\def\LNU{L(\NU )}
\def\Lin#1{L^{-1}(#1)}
\def\L#1{L(#1)}
\def\Li#1{L_i(#1)}
\def\FNU#1{F_\NU(#1)}
\def\FinNU#1{F^{-1}_\NU(#1)}

\def\ftthetax{\tilde{f}_{\Theta|X}(\theta|x)}
\def\ER{R}

\def\Ndd#1#2#3{{\cal N}\left( #1 ;~ #2, #3 \right)}
\def\Cdd#1#2#3{{\cal C}\left( #1 ;~ #2, #3 \right)}
\def\Sdd#1#2#3#4{{\cal S}\left( #1 ;~ #2, #3, #4 \right)}
\def\Edd#1#2{{\cal E}\left( #1 ;~ #2 \right)}
\def\DEdd#1#2{{\cal DE}\left( #1 ;~ #2 \right)}
\def\Gdd#1#2#3{{\cal G}\left( #1 ;~ #2, #3 \right)}
\def\IGdd#1#2#3{{\cal IG}\left( #1 ;~ #2, #3 \right)}

\def\Xwr{X(\omega,\rb)}
\def\xwr{x(\omega,\rb)}

\def\muzw{\mu_{z}(\omega)}
\def\muzwr{\mu_{z(\rb)}(\omega)}
\def\szw{\sigma_{z}^2(\omega)}
\def\szwr{\sigma_{z(\rb)}^2(\omega)}
\def\pzw{p_{z}(\omega)}
\def\pzwr{p_{z(\rb)}(\omega)}

\def\hszw{\widehat{\sigma_{z}^2}(\omega)}
\def\hmuzw{\widehat{\mu}_{z}(\omega)}
\def\hpzw{\widehat{p}_{z}(\omega)}

\def\muzwrl{\mu_{z(\rb)}(\omega-l)}

\def\pxwrcz{p\left(\xwr ~|~ z\right)}
\def\pxwr{p\left(\xwr\right)}

\def\Xbw{\Xb(\omega)}
\def\xbw{\xb(\omega)}
\def\uXb{\underline{\Xb}}
\def\uxb{\underline{\xb}}
\def\pxbw{p(\xbw)}
\def\puxb{p(\uxb)}
\def\pzb{P(\zb)}

\def\Szw{S_z(\omega)}
\def\mubw{\mub(\omega)}

\def\Srw{S_{\rb}(\omega)}
\def\Swr{S_{\omega}(\rb)}
\def\Sbw{\Sb(\omega)}
\def\sbw{\sb(\omega)}
\def\Sbr{\Sb(\rb)}
\def\mubw{\mub(\omega)}
\def\cov#1{\mbox{cov}[#1]}

\def\pdf{\emph{pdf}}
\def\pmf{\emph{pmf}}
\def\dpdx#1#2{\frac{\partial #1}{\partial #2}}

\def\REM#1{}

\def\XXwr{X(\omega,\rb)}
\def\xxwr{x(\omega,\rb)}

\def\Xwr{X_{\omega}(\rb)}
\def\xwr{x_{\omega}(\rb)}
\def\Xrw{X_{\rb}(\omega)}
\def\xrw{x_{\rb}(\omega)}

\def\Ywr{Y_{\omega}(\rb)}
\def\ywr{y_{\omega}(\rb)}
\def\Yrw{Y_{\rb}(\omega)}
\def\yrw{y_{\rb}(\omega)}

\def\Ewr{E_{\omega}(\rb)}
\def\ewr{\epsilon_{\omega}(\rb)}
\def\Erw{E_{\rb}(\omega)}
\def\erw{\epsilon_{\rb}(\omega)}

\def\Xbr{\Xb(\rb)}
\def\xbr{\xb(\rb)}
\def\Xbw{\Xb(\omega)}
\def\xbw{\xb(\omega)}

\def\uXb{\underline{\Xb}}
\def\uxb{\underline{\xb}}
\def\uXbz{\underline{\Xb}_z}
\def\uxbz{\underline{\xb}_z}

\def\uthetab{\underline{\thetab}}
\def\uthetabz{\underline{\thetab}_z}

\def\Ywr{Y_{\omega}(\rb)}
\def\ywr{y_{\omega}(\rb)}
\def\Yrw{Y_{\rb}(\omega)}
\def\yrw{y_{\rb}(\omega)}
\def\Ybr{\Yb(\rb)}
\def\ybr{\yb(\rb)}
\def\Ybw{\Yb(\omega)}
\def\ybw{\yb(\omega)}
\def\uYb{\underline{\Yb}}
\def\uyb{\underline{\yb}}
\def\uYbz{\underline{\Yb}_z}
\def\uybz{\underline{\yb}_z}

\def\Ewr{E_{\omega}(\rb)}
\def\ewr{\epsilon_{\omega}(\rb)}
\def\Erw{E_{\rb}(\omega)}
\def\erw{\epsilon_{\rb}(\omega)}
\def\Ebr{Eb(\rb)}
\def\ebr{\epsilonb(\rb)}
\def\Ebw{Eb(\omega)}
\def\ebw{\epsilonb(\omega)}
\def\uEb{\underline{\Eb}}

\def\Erwp{E_{\rb}(\omega')}
\def\Xrwp{X_{\rb}(\omega')}

\def\xwmr{x_{\omega-1}(\rb)}
\def\muzwm{\mu_{z}(\omega-1)}

\def\rhoz{\rho_z}
\def\sez{\sigma_{\eta_z}^2}

\def\xwrp{x_{\omega}(\rb')}
\def\xrwp{x_{\rb}(\omega')}
\def\xwpr{x_{\omega'}(\rb)}
\def\xwprp{x_{\omega'}(\rb')}
\def\Xwrp{x_{\omega}(\rb')}
\def\Xwrp{X_{\omega}(\rb')}
\def\Xwpr{X_{\omega'}(\rb)}
\def\Xwprp{X_{\omega'}(\rb')}

\def\sigmaed{\sigmae^2}
\def\sigmaew{\sigmae(\omega)}
\def\sigmaedw{\sigmaed(\omega)}

\def\Sigmabe{\Sigmab_{\epsilon}}
\def\sigmabh{\widehat{\Sigmab}}
\def\xbh{\widehat{\xb}}
\def\zbh{\widehat{\zb}}
\def\qbh{\widehat{\qb}}
\def\sbh{\widehat{\sb}}

\def\argmax#1#2{\arg\max_{#1}\left\{#2\right\}}
\def\argmin#1#2{\arg\min_{#1}\left\{#2\right\}}
\def\espx#1#2{\mbox{E}_{#1}\left\{#2\right\}}
\def\esp#1{\mbox{E}\left\{#1\right\}}
\def\d#1{\mbox{~d}#1}

\def\Tr#1{\mbox{Tr}\left[#1\right]}

\def\pmatrix#1{\left[\barr{c} #1 \earr\right]}
\def\Bf{\bm{B}}
\def\gbh{\widehat{\gb}}
\def\dbh{\widehat{\db}}
\def\Tr#1{\mbox{Tr}\left[#1\right]}
\def\Cov#1{\mbox{Cov}\left[#1\right]}
\def\iid{i.i.d.~}

\def\cro#1{\left[#1\right]}
\def\acc#1{\left\{#1\right\}}
\def\aprio{\emph{a priori~}}
\def\apost{\emph{a posteriori~}}
\def\hbh{\widehat{\hb}}
\def\qh{\widehat{q}}
\def\fbt{\widetilde{\fb}}
\def\thetabt{\widetilde{\thetab}}